\documentclass[noeprint,noshowpacs,longbibliography,nopreprintnumbers,twocolumn,prb,showpacs,superscriptaddress,amsmath,amssymb,citeautoscript,10pt,aps]{revtex4-2}
\usepackage{graphicx}
\usepackage{dcolumn}
\usepackage{color}
\usepackage{bm}
\usepackage[hidelinks]{hyperref}
\hypersetup{
    colorlinks,
    citecolor=blue,
    filecolor=blue,
    linkcolor=blue,
    urlcolor=blue
}
\graphicspath{{Figures/}{./Figures/}{.}}

\DeclareMathOperator{\sgn}{sgn}
\DeclareMathOperator{\Tr}{Tr}
\DeclareMathOperator{\C}{\mathcal{C}}
\DeclareMathOperator{\CL}{\mathcal{L}}
\newcommand{\bsigma}{{\bm \sigma}}
\newcommand{\btau}{{\bm \tau}}

\begin{document}

\title{Quantized Thermal Hall Conductance and the Topological Phase Diagram of a Superconducting Bismuth Bilayer}

\author{Szczepan G{\l}odzik}
\affiliation{Institute of Physics, M.~Curie-Sk{\l}odowska University, 20-031 Lublin, Poland}
\affiliation{Jo\v{z}ef \v{S}tefan Institute, Jamova 39, SI-1000 Ljubljana, Slovenia}
\author{Nicholas Sedlmayr}
\email{sedlmayr@umcs.pl}
\affiliation{Institute of Physics, M.~Curie-Sk{\l}odowska University, 20-031 Lublin, Poland}

\date{\today}

\begin{abstract}
Two dimensional topological superconductors with chiral edge modes are predicted to posses a quantized thermal Hall effect proportional to the Chern number, exactly half that for chiral topological insulators. However not much work has been done in identifying the quantized heat conductance in the literature, even for some of the standard models of topological superconductivity. Here we introduce a model based on a proximity induced superconducting Bismuth bilayer, and directly calculate the thermal Hall conductance of this lattice model. This model serves as a demonstration of the state of the art possible in such a calculation, as well as introducing an interesting paradigmatic topological superconductor with a rich phase diagram. We demonstrate the quantized thermal Hall plateaus in several different topological phases, and compare this to numerical calculations of the Chern number, as well as analytical calculations of the Chern number's parity invariant. We demonstrate that it is possible to get a reasonable topological phase diagram from the quantized thermal Hall calculations. The technique used can be applied to wide range of models directly in real space.
\end{abstract}

\maketitle

\section{Introduction}

Applying a thermal gradient across an appropriate sample can lead to a non-dissipative thermal Hall current, a phenomenon known as the Leduc-Righi effect~\cite{Pitaevskii2012}. Furthermore in fully gapped systems with chiral edge modes this results in a non-zero quantized thermal Hall conductance $\kappa/T$~\cite{Kane1997,Read2000}. In fractional quantum Hall states it can be shown that the thermal Hall conductance is quantized to $c\pi/6$ in units of $ k_B^2/\hbar$ where $c$ is the central charge of the conformal field theory describing the edge modes, whereas for a non-interacting topological superconductor one has $c=\nu/2$, where $\nu$ is the Chern number, leading to $\kappa/T=\nu\pi/12$ in units of $ k_B^2/\hbar$~\cite{Kane1997,Read2000,Vishwanath2001,Sumiyoshi2013}. In view of this relation between the thermal Hall effect and the Chern number the quantized thermal Hall conductance is a potentially very useful probe of the topology of Chern insulators and superconductors~\cite{Kane1997,Read2000,Vishwanath2001,Cappelli2002,Qin2011,Sumiyoshi2013,Shimizu2015,Metavitsiadis2017a,Tang2019a,Ngampruetikorn2020,Fulga2020,Yang2020,Yang2022}.

Two-dimensional topological superconductors themselves have also seen an explosion of interest, for some early examples see Refs.~\onlinecite{Fu2008,Potter2010,Mizushima2013,Wang2014,Poyhonen2014,Seroussi2014,Wakatsuki2014,Deng2014,San-Jose2014,Thakurathi2014,Mohanta2014,Bjornson2015,Sedlmayr2015,Rontynen2016,Kaladzhyan2017,Yang2016,Sedlmayr2017,Kezilebieke2022}. However, despite the relation between these systems and the quantized thermal Hall effect being well established, there are relatively few explicit calculations of the thermal current or conductance in these systems. Some exceptions include direct numerical calculation of the thermal Hall conductance for simple tight binding Hamiltonians~\cite{Tang2019a}, the linearized regime of mixed state $d$-wave superconductors~\cite{Vishwanath2001}, and direct calculations of the thermal Hall effect from the system bulk~\cite{Sumiyoshi2013} or from a generalized Wiedemann-Franz law~\cite{Shimizu2015}. Here we show that lattice calculations can be performed for even fairly complicated models directly in real space to calculate the thermal Hall conductance. Thermal currents have also been considered in the context of two-dimensional $Z_{2}$ spin liquids~\cite{Metavitsiadis2017a}, as a probe of superconducting gap anisotropy~\cite{Hu2022}, and in an interacting Kitaev-Heisenberg model~\cite{Kumar2023}.

On the experimental side the situation is rather complicated due to the subtleties in measuring the required thermal currents. Nonetheless impressive progress has been made along this direction looking at fractional quantum Hall states~\cite{Srivastav2022}, the topological thermal Hall conductance in gallium arsenide based topological insulators~\cite{Melcer2023}, and in a graphene aerogel~\cite{Cox2023}. Another experiment has studied heat dissipation to determine the thermal decay length of the edge states~\cite{Moore2023}. A relatively recent experiment measured an unusually high value of the thermal Hall conductivity in the pseudogap phase of a few cuprates~\cite{Grissonnanche2019}, an effect explained theoretically as the result of the orbital coupling driving the system close to a chiral spin liquid phase with spinons becoming the neutral carriers responsible for the thermal conductance~\cite{Samajdar2019}. Using noise thermometry, the quantum limit for thermal transport across a single channel has been measured in quantum point contacts in two dimensional electron gases~\cite{Jezouin2013} and in graphene based devices~\cite{Srivastav2019a}. Thermal conductance quantization has also been measured in fractional quantum hall states~\cite{Banerjee2017,Srivastav2021,Dutta2022,LeBreton2022,Melcer2022a,Hein2023}. Finally we note that phonons may alter the perfect quantization leading to an only approximately quantized thermal conductance~\cite{Banerjee2018a,Vinkler-Aviv2018}.

In a two-dimensional topological superconductor, the edge states are often composed of electrically neutral Majorana fermions. As Majorana fermions are each half an electron this is often understood as the reason for the factor of one half difference between the quantized thermal Hall conductance for the chiral insulators and superconductors. However it should be noted that in fact not all chiral bands inside the gap are composed of Majorana fermions. Only those bands which pass through the time reversal invariant momenta can be Majorana fermions, and this number can even depend on the orientation of the edge~\cite{Sedlmayr2017}. The Chern number predicts the number of topologically protected bands crossing the gap, but not the number of Majorana fermion bands, the only additional piece of information one can infer is that for an odd Chern number the edge must possess at least one band of Majorana fermions but could in principle have more. In our results we find that the quantized thermal Hall effect is always in agreement with the number of chiral bands, and hence the Chern number, and not the number of Majorana fermions. This result is in agreement with calculations performed entirely using bulk states~\cite{Sumiyoshi2013}.

Additionally two-dimensional topological superconductors have been investigated with other standard techniques such as scanning tunneling microscopy and spectroscopy~\cite{Menard2015,Menard2017,Menard2019a,Kezilebieke2020,Bazarnik2023,Wong2023a}. In Josephson junction setups supercurrents have been measured~\cite{Kurter2014,Stehno2016}, interference measurements have been done~\cite{Kurter2014a}, as well as spectroscopy~\cite{Fornieri2019,Ren2019}. A direct measurement of the topological index remains however elusive.

The model which we introduce here is based on bismuth~\cite{Hofmann2006,Schindler2018a,Aguilera2021}, a common component of topological materials due to its strong Rashba spin orbit coupling~\cite{Koroteev2004,Ohtsubo2012,Takayama2015}, and an interesting material in its own right. Edge states are known to exist in 3, 5, and 7 monolayer thick bismuth structures~\cite{Salehitaleghani2023}, though for a different structure than the one we will consider. We focus on a (111) bilayer bismuth structure, a topological insulator which possesses edge modes~\cite{Wada2011,Kotaka2012,Yang2012,Sabater2013,Drozdov2014,Lima2015,Niu2015,Demidov2020}. The structure of the (111) surface of bismuth is known from a low-energy electron diffraction (LEED) analysis and first-principles calculations~\cite{Monig2005,Koroteev2008}. The (111) bismuth bilayer we consider is an AB stacked and warped hexagonal lattice with strongest inter-layer coupling between the non-stacked BA atoms~\cite{Hofmann2006}. This bilayer then has proximity induced $s$-wave superconductivity from a substrate. The bulk topological phase diagram for isotropic and non-isotropic hexagonal lattices with spin-orbit coupling has been extensively studied~\cite{Black-Schaffer2014,Dutreix2014,Sedlmayr2015,Wang2016,Dutreix2017,Sedlmayr2017,Kaladzhyan2017,Pangburn2023,Crepieux2023,Pangburn2023a}, however the model we introduce here has not been previously studied and we first calculate its topological properties and Chern number. This model proves a fertile playground for considering the thermal Hall conductance both due to its complexity, which probes the limits of the techniques we use, but also because it has a rich phase diagram with relatively large Chern numbers possible. This allows us to test the quantized thermal Hall conductance in many different topological phases.

Although it has been shown that for non-interacting fermionic tight binding Hamiltonians it is possible to numerically calculate the thermal Hall conductance~\cite{Tang2019a}, the limits of applicability of this approach have not been tested. Here we probe the state of the art for such calculations by taking a relatively complicated model with calculations purely on a real space lattice, which we find converges faster, and performing numerical calculations for the largest lattices we are able to. The practical issues of calculation of sometimes very complicated commutators we partially solve with new software~\cite{Zitko2011a}. The thermal Hall conductance is compared directly to the Chern number to test its quantization. As the calculation of the Chern number is also numerically demanding we also perform analytical calculations of the parity of the Chern number, which allows for exact results. These are compared to the band structures showing the edge modes along both zig-zag and armchair edges of the bilayer. Overall we demonstrate here how to apply these techniques to a very wide range of topological superconductor models.

This article is organised as follows. In Sec.~\ref{sec_model} we introduce the tight binding model describing the proximitized Bismuth bilayer on which we base our calculations. In Sec.~\ref{sec_topology} we demonstrate the analytical calculations of the Chern number parity, as well as the numerical calculations of the Chern number, and their results. Sec.~\ref{sec:kappa} compares this to the direct calculations of the thermal Hall conductance. In Sec.~\ref{sec:con} we conclude with a discussion of the results.

\section{The Tight Binding Model}\label{sec_model}

\begin{figure}
\includegraphics*[width=0.85\columnwidth,keepaspectratio]{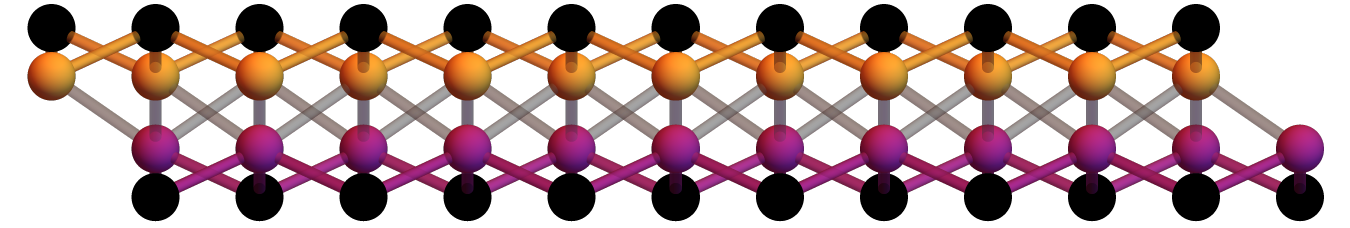}\\
\includegraphics*[width=0.85\columnwidth,keepaspectratio]{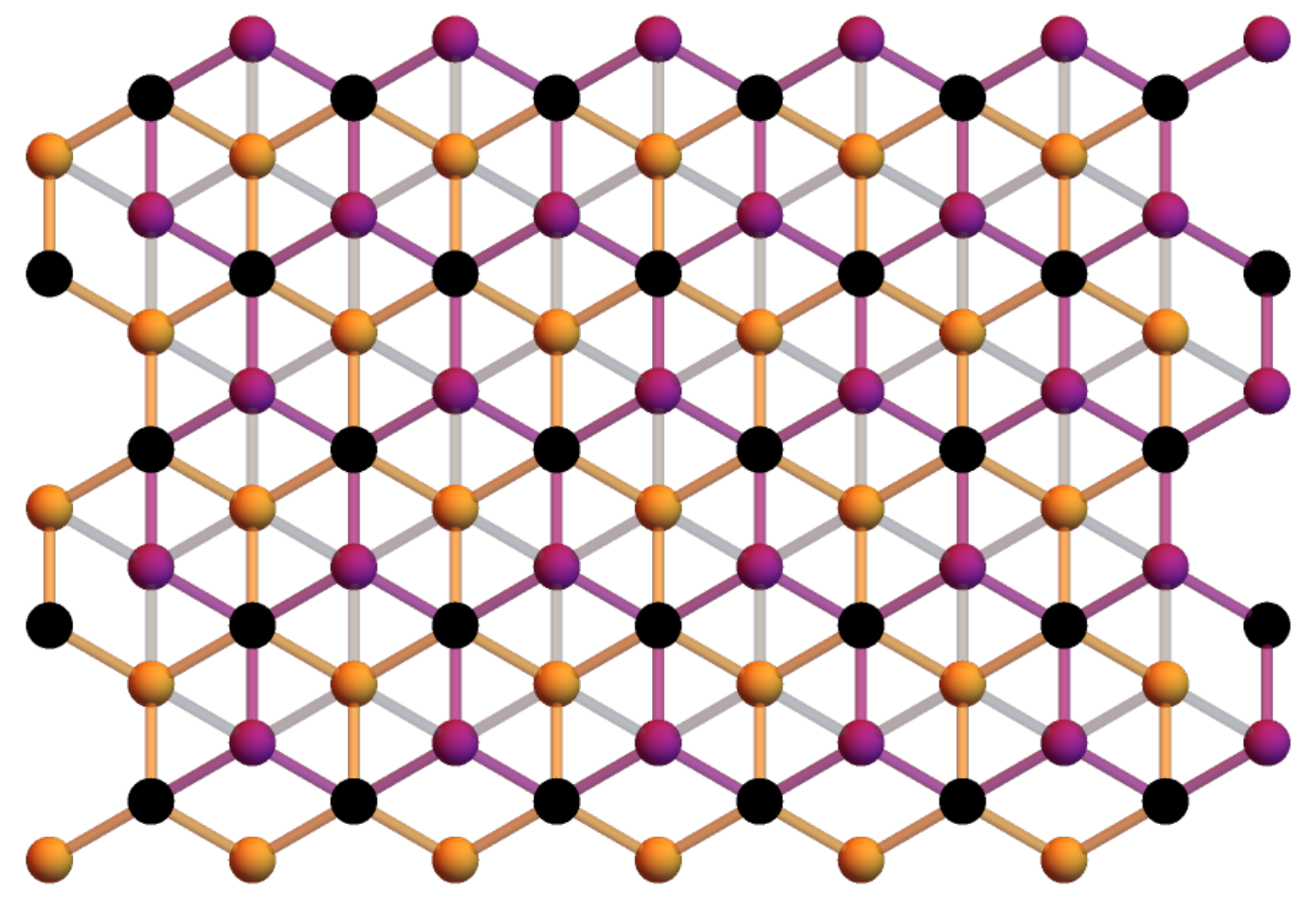}
\caption{(Color online) Sideways on and top down schematics of the bilayer system for Bismuth (111). The upper layer consists of black (A) and yellow (B) atoms. The lower layer consists of black (B) and purple (A) atoms. Gray lines show the inter layer hopping terms. The A atoms in the upper and B atoms in the lower plane (both black) are not coupled due to the buckling of each layer~\cite{Hofmann2006}.
}
\label{ab_schematic}
\end{figure}

We start from a Bogoliubov--de-Gennes (BdG) Hamiltonian for the particle and hole spaces on a general two dimensional lattice written in the Nambu basis, $\hat\Psi_{j}=(\hat{c}_{j,\uparrow},\hat{c}_{j,\downarrow},\hat{c}^{\dag}_{j,\downarrow},-\hat{c}^{\dag}_{j,\uparrow})^T$, where $\hat{c}^{\dag}_{j,\sigma}$ creates a particle of spin $\sigma$ at site $j$. We use Pauli matrices $\vec{\bsigma}$ for the spin subspace and $\vec{\btau}$ for the particle-hole subspace, the full space is therefore a tensor product between these spaces. However we will adopt the common notation that this is written as  ${\btau}^a\otimes{\bsigma}^b\to{\btau}^a{\bsigma}^b$ and identity matrices are not written at all.  The lattice structure is based upon a bismuth (111) bilayer~\cite{Hofmann2006}, which consists of two AB stacked hexagonal lattices, with warping in the vertical direction, see Fig.~\ref{ab_schematic}. Due to the warping there is no hopping between the vertically aligned A and B sites, but there is inter-layer hopping between adjacently spaced sites. Both inter- and intra-layer hopping terms form hexagonal lattices. Note that we set $t=\hbar=k_B=1$ throughout. 

Crucially the Hamiltonians we consider anti-commute with an anti-unitary particle hole operator, $\{\C,H\}=0$, where $\C=e^{ i\varphi}{\btau}^y{\bsigma}^yK$ with $K$ complex conjugation and $\varphi$ an arbitrary phase. There is no chiral or time reversal symmetry present which means this system is in the D class of the topological periodic table and has a $\mathbb{Z}$ invariant~\cite{Schnyder2009}, which is here the Chern number~\cite{Thouless1982,Bernevig2013}. We calculate both the Chern number and its parity~\cite{Sato2009} in Sec.~\ref{sec_topology}.

The full lattice Hamiltonian, written in BdG form, is
\begin{equation}\label{hamiltonian}
    \hat{\mathcal{H}}= \hat{\mathcal{H}}_0+\hat{\mathcal{H}}_1+ \hat{\mathcal{H}}_{\rm IL}.
\end{equation}
The first contribution consist of the onsite terms:
\begin{equation}\label{onsite}
    \hat{\mathcal{H}}_0=-\sum_{j}\hat\Psi^\dagger_{j}
    \left[\mu{\btau}^z+\Delta{\btau}^x+B{\bsigma}^z\right]\hat\Psi_{j}.
\end{equation}
$\mu$ is the chemical potential, $B$ a Zeeman field, and $\Delta$ a proximity induced s-wave superconducting pairing. The second contribution consists of the intra-plane hopping terms:
\begin{equation}\label{hopping}
    \hat{\mathcal{H}}_1=-
    \sum_{\langle j,\ell\rangle }\hat\Psi^\dagger_{j}
    \left[t{\btau}^z
    -i\alpha  {\btau}^z\hat{z}\cdot\left(\vec{\delta}_{j\ell}\times{\vec{\bsigma}}\right)\right]\hat\Psi_{\ell}.
\end{equation}
$\alpha$ is a Rashba spin-orbit interaction, present when inversion symmetry is broken, and $t$ is the usual hopping strength. $\langle j,\ell\rangle$ is used to refer to nearest neighbour pairs of sites $j$ and $\ell$ in each layer, but not between the layers. $\vec{\delta}_{j\ell}$ is the real space vector between sites $j$ and $\ell$. Finally the inter layer hopping is
\begin{equation}\label{inter_layer}
\hat{\mathcal{H}}_{\rm IL}=-t'\sum_{\langle j,\ell\rangle' }\hat\Psi^\dagger_{j}{\btau}^z\hat\Psi_{\ell}\,.
\end{equation}
$\langle j,\ell\rangle'$ denotes pairs of sites in different layers, connected by the hopping terms shown as gray lines in Fig.~\ref{ab_schematic}.

After a standard Fourier transform the Hamiltonian can be written as $\hat{\mathcal{H}}=\sum_{\vec k}\hat\Psi^\dagger_{\vec k}\mathcal{H}(\vec k)\hat\Psi_{\vec k}$ with
\begin{equation}\label{kham}
\mathcal{H}(\vec k)=\begin{pmatrix}
{\bm f}_{\vec k}-B & \CL_{\vec k} & -\Delta & 0\\
\CL^\dagger_{\vec k}& {\bm f}_{\vec k}+B & 0 & -\Delta\\
 -\Delta  &0 & -B-{\bm f}^*_{-\vec k} & \CL^T_{-\vec k} \\
0&  -\Delta &  \CL^*_{-\vec k}& -{\bm f}^*_{-\vec k}+B
\end{pmatrix}.
\end{equation}
Each entry in this matrix is itself a matrix for the sublattice structure. Labelling the upper and lower layers as $U$ and $L$, and the sublattice sites in each layer as $A$ and $B$, the explicit form of the operator is
\begin{equation}
    \hat\Psi^T_{\vec k}=
    \begin{pmatrix}\hat\psi_{k,\uparrow} &\hat\psi_{k,\downarrow}
    &\hat\psi^\dagger_{k,\downarrow}
    &-\hat\psi^\dagger_{k,\uparrow}
    \end{pmatrix}
\end{equation}
with
\begin{equation}
    \hat\psi^T_{\vec k,\sigma}=
    \begin{pmatrix}\hat c_{\vec k,\sigma,UA} &\hat c_{\vec k,\sigma,UB} &\hat c_{\vec k,\sigma,LA} &\hat c_{\vec k,\sigma,LB}
    \end{pmatrix}.
\end{equation}
The sublattice matrices are
\begin{equation}
{\bm f}_{\vec k}=-\sum_{j=1}^3\begin{pmatrix}
\mu/3 & te^{ i \vec k\cdot\vec\delta_j} & 0 & 0\\
 te^{- i \vec k\cdot\vec\delta_j}& \mu/3  &  t'e^{- i \vec k\cdot\vec\delta_j} & 0\\
 0 &  t'e^{ i \vec k\cdot\vec\delta_j} & \mu/3  & te^{ i \vec k\cdot\vec\delta_j}\\
  0 & 0 & te^{- i \vec k\cdot\vec\delta_j}& \mu/3 
\end{pmatrix}
\end{equation}
and
\begin{equation}
\CL_{\vec k}=\alpha\sum_{j=1}^3\vec\delta_j\cdot\left(1, i\right)\begin{pmatrix}
0 & -e^{ i \vec k\cdot\vec\delta_j} & 0 & 0\\
 e^{- i \vec k\cdot\vec\delta_j}& 0 & 0 & 0\\
 0 & 0 & 0&-e^{ i \vec k\cdot\vec\delta_j} \\
0 & 0& e^{- i \vec k\cdot\vec\delta_j}& 0
\end{pmatrix}.
\end{equation}
Both $B$ and $\Delta$ are diagonal in the sublattice space. 
$\vec\delta_j$ are the nearest neighbour vectors between A and B atoms
\begin{equation}\label{deltavector}
\{\vec\delta_1,\vec\delta_2,\vec\delta_3\}=\left\{\left(\frac{\sqrt{3}}{2},-\frac{1}{2}\right),\left(-\frac{\sqrt{3}}{2},-\frac{1}{2}\right),\left(0,1\right)\right\}.
\end{equation}
See Fig.~\ref{ab_schematic} for a schematic of the convention used. For completeness we finally note that
\begin{equation}
\vec k=\left(\frac{2\pi n}{\sqrt{3}N_x},\frac{4\pi m}{3N_y}\right)\,,
\end{equation}
where $n=1,2,\ldots N_x$ and $m=1,2,\ldots N_y$. In this form the time reversal invariant (TRI) momenta are $\vec\Gamma_i=\{(0,0),(0,2\pi/3),(\pi/\sqrt{3},\pi/3),(\pi/\sqrt{3},-\pi/3)\}$. Throughout the rest of this paper we focus on this model.

In table \ref{tab:par} we give the parameters which are used for the different examples contained throughout the paper when considering bandstructures, thermal Hall conductance, and current maps. We demonstrate our results with examples from several different topological phases and focus on parameter ranges which allow us to find reasonable numerical results at the largest system sizes we are able to probe. 

\begin{table}
    \centering
    \begin{tabular}{|c|c|c|c|c|c|}\hline
    \quad$\nu$\quad{} & \quad$\mu$\quad{} & \quad$B$\quad{} & \quad$\alpha$\quad{} & $\quad\Delta$\quad{} & \quad$t'$\quad{} \\\hline
       0  & $1.2t$ & $2t$ & $0.3t$ & $0.4t$ & $0.5t$ \\\hline
       -1  & $0.8t$ & $3.5t$ & $0.3t$ & $0.4t$ & $0.5t$ \\\hline
       2  & $3.1t$ & $1.3t$ & $0.3t$ & $0.4t$ & $0.5t$ \\\hline
       4  & $2t$ & $2.3t$ & $0.3t$ & $0.4t$ & $0.5t$ \\\hline
       -5  & $1.6t$ & $1.2t$ & $0.3t$ & $0.4t$ & $0.5t$ \\\hline
    \end{tabular}
    \caption{Parameters for the examples used throughout this paper. Examples are taken from phases with several different Chern numbers $\nu$.}
    \label{tab:par}
\end{table}

\begin{figure}
\includegraphics*[width=0.475\columnwidth]{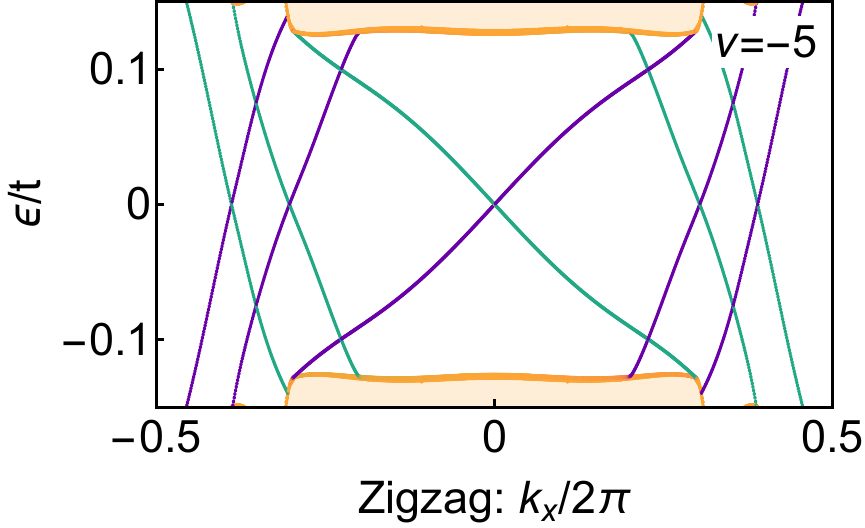}
\includegraphics*[width=0.475\columnwidth]{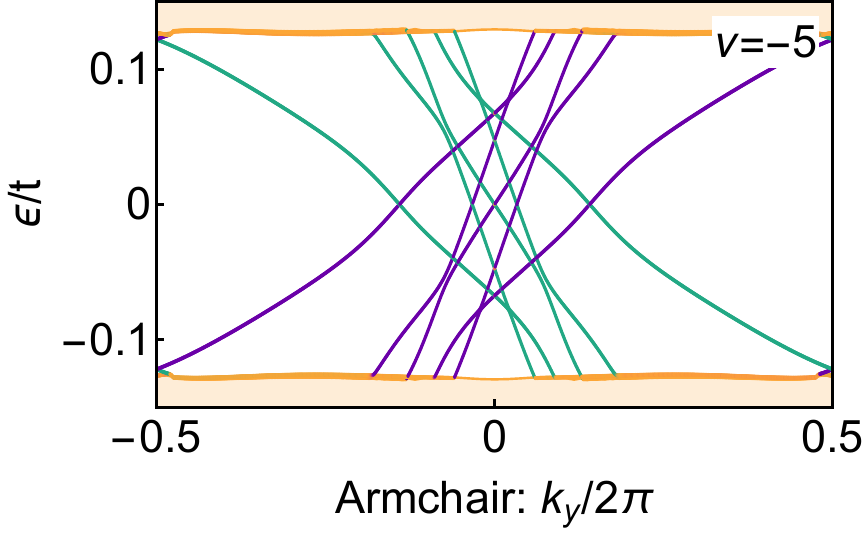}
\caption{(Color online) Band structures projected along the $x$ and $y$ directions for zig-zag, panel (a), and armchair, panel (b), edges respectively for a point in the $\nu=-5$ phase, see Fig.~\ref{fig:phase_1}. Parameters are given in table \ref{tab:par}. Light yellow shows the bulk bands, and the green and purple lines show the edge modes on the two different edges. Along the edge a single mode passes through $k_x=0$, which is a Majorana fermion mode, all other modes are not Majorana fermions~\cite{Sedlmayr2017,Beenakker2020}.}
\label{ba_bilayer_band_ex}
\end{figure}

\begin{figure}
\includegraphics*[width=0.475\columnwidth]{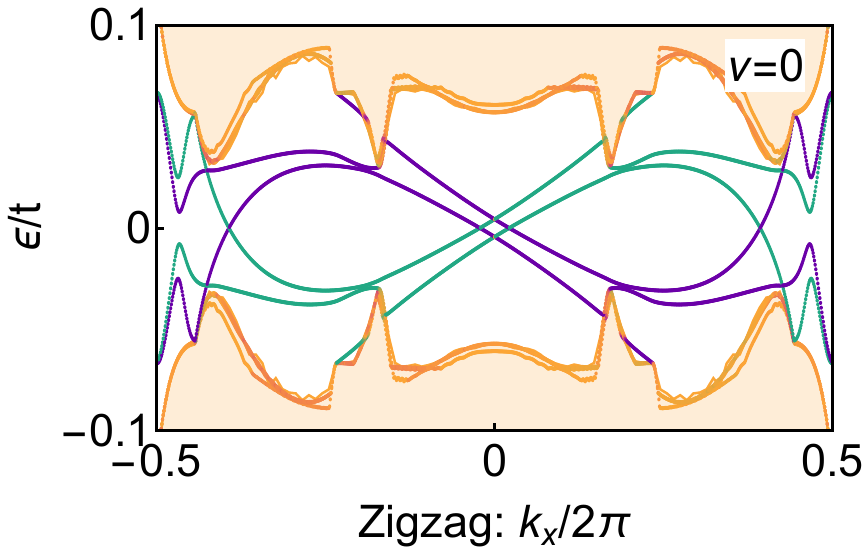}
\includegraphics*[width=0.475\columnwidth]{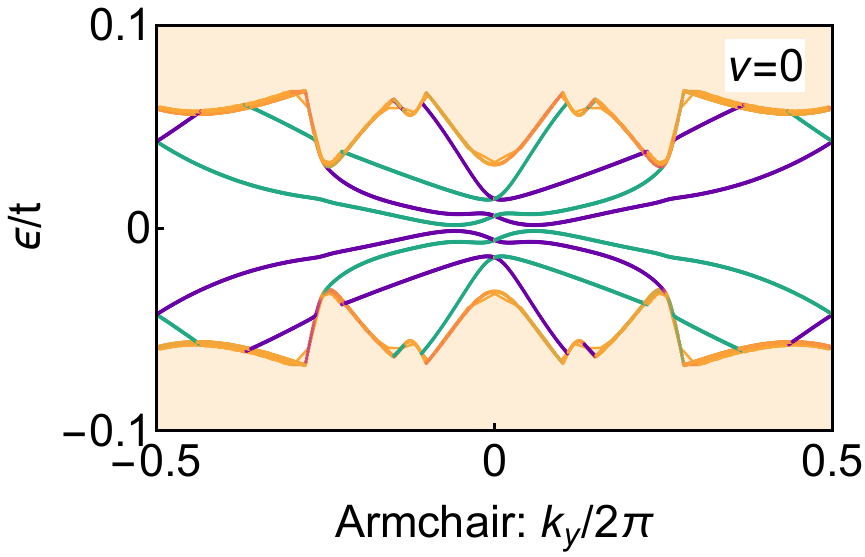}
\caption{(Color online) Band structures projected along the $x$ and $y$ directions for zig-zag, panel (a), and armchair, panel (b), edges respectively for a point in the $\nu=0$ phase, see Fig.~\ref{fig:phase_1}. Parameters are given in table \ref{tab:par}. Light yellow shows the bulk bands, and the green and purple lines show the edge modes on the two different edges. A careful view will show that there are no topologically protected modes, in agreement with the bulk boundary correspondence.}
\label{ba_bilayer_band_ex2}
\end{figure}

Examples for the bandstructure showing the edge modes along both zig-zag and armchair edges are shown in Figs.~\ref{ba_bilayer_band_ex} and \ref{ba_bilayer_band_ex2}. From the bulk boundary correspondence we know that the Chern number is equal to the number of topologically protected bands on an edge. Some further examples are given in appendix \ref{app:more_bands} for the other exemplary points in the phase diagram we use. We take the Hamiltonian given by Eq.~\eqref{hamiltonian} and Fourier transform along either the $x$ direction for zig-zag edges, or along the $y$ direction for armchair edges. The perpendicular direction is kept finite with open boundary conditions, resulting in edge modes propagating along the two edges.

In Fig.~\ref{ba_bilayer_band_ex2} we show the bandstructure in the $\nu=0$ phase. According to the bulk boundary correspondence there should be no topologically protected edge modes. A close look reveals that although there are edge modes present, they are not topologically protected as they always connect the upper bulk bands to themselves or the lower bulk bands to themselves, and can therefore easily be removed by a continuous deformation. This $\nu=0$ phase therefore gives a nice check for the quantized Hall conductance as it should result in zero, despite the presence of these edge modes.

\section{Topological Phase diagram}\label{sec_topology}

One of the main focuses of this article is the quantized thermal Hall conductance, which is proportional to the Chern number, $\nu\in\mathbb{Z}$. Hence we wish to compare the calculated thermal conductance and an {\it independent} calculation of the Chern number. Naturally the calculation of the Chern number also informs us about the rich topological phase diagram of this model. The Chern number for such a complicated model can only be calculated numerically, however we can find an analytical expression for its parity $\delta=(-1)^\nu$. Both of these can be compared to the band structure, as one expects that the Chern invariant gives the number of chiral propagating modes along the edge of the system due to the bulk boundary correspondence. We remind the reader here that the chiral modes are not necessarily Majorana fermion modes~\cite{Beenakker2020} and the number of chiral modes which {\it are} Majorana fermion modes can depend on the type of edge~\cite{Sedlmayr2017}.

\subsection{Chern Number}

The Chern number, equivalent to the TKNN invariant~\cite{Thouless1982}, can be calculated numerically~\cite{Ghosh2010,Wang2016,Sedlmayr2017}. It is given by
\begin{align}\label{tknn}
    \nu=i\int \frac{d^2kd\omega}{8\pi^2} \Tr&\bigg[G_{\vec k,\omega}^2\left[\partial_{k_y}\mathcal{H}(\vec k)\right]G_{\vec k,\omega}\left[\partial_{k_x}\mathcal{H}(\vec k)\right]
    \nonumber\\&-G_{\vec k,\omega}^2\left[\partial_{k_x}\mathcal{H}(\vec k)\right]G_{\vec k,\omega}\left[\partial_{k_y}\mathcal{H}(\vec k)\right]\bigg]\,,
\end{align}
with the Green's function $G_{\vec k,\omega}=(\mathcal{H}(\vec k)- i\omega)^{-1}$ and momentum $\vec{k}=(k_x,k_y)$. By diagonalizing the Hamiltonian the frequency integral in Eq.~\eqref{tknn} can be performed analytically. Let $S$ be a rotation to the energy eigenbasis of the Hamiltonian so that $S^{-1}\mathcal{H}(\vec k)S$ is diagonal with eigenvalues $\varepsilon_j$. If we define $S^{-1}\partial_{k_{x,y}}\mathcal{H}(\vec k)S\equiv\mathcal{H}^{x,y}$ then we find
\begin{equation}
\nu =i\int \frac{d^2kd\omega}{8\pi^2} \sum_{j \ell}
\frac{\mathcal{H}^{y}_{j \ell}\mathcal{H}^{x}_{\ell j}-\mathcal{H}^{x}_{j \ell}\mathcal{H}^{y}_{\ell j}}{(\varepsilon_j- i\omega)^2(\varepsilon_\ell- i\omega)}\,.
\end{equation}
Performing the frequency integral leaves~\cite{Thouless1982}
\begin{equation}
\nu=\int \frac{d^2k}{2\pi} \sum_{j \ell}\mathcal{H}^{x}_{j \ell}\mathcal{H}^{y}_{\ell j}\frac{\sgn[\varepsilon_j]\Theta[-\varepsilon_j\varepsilon_\ell]}{(\varepsilon_j-\varepsilon_\ell)^2}\,,
\end{equation}
where $\Theta$ is the Heaviside theta function. The momentum integral over the Brillouin zone can be performed numerically, with the matrix multiplication and determination of $S$ also calculated numerically. For the integration we use Romberg's method over the hexagonal Brillouin zone~\cite{Suli2003}.

\begin{figure*}
\includegraphics[height=0.6\columnwidth]{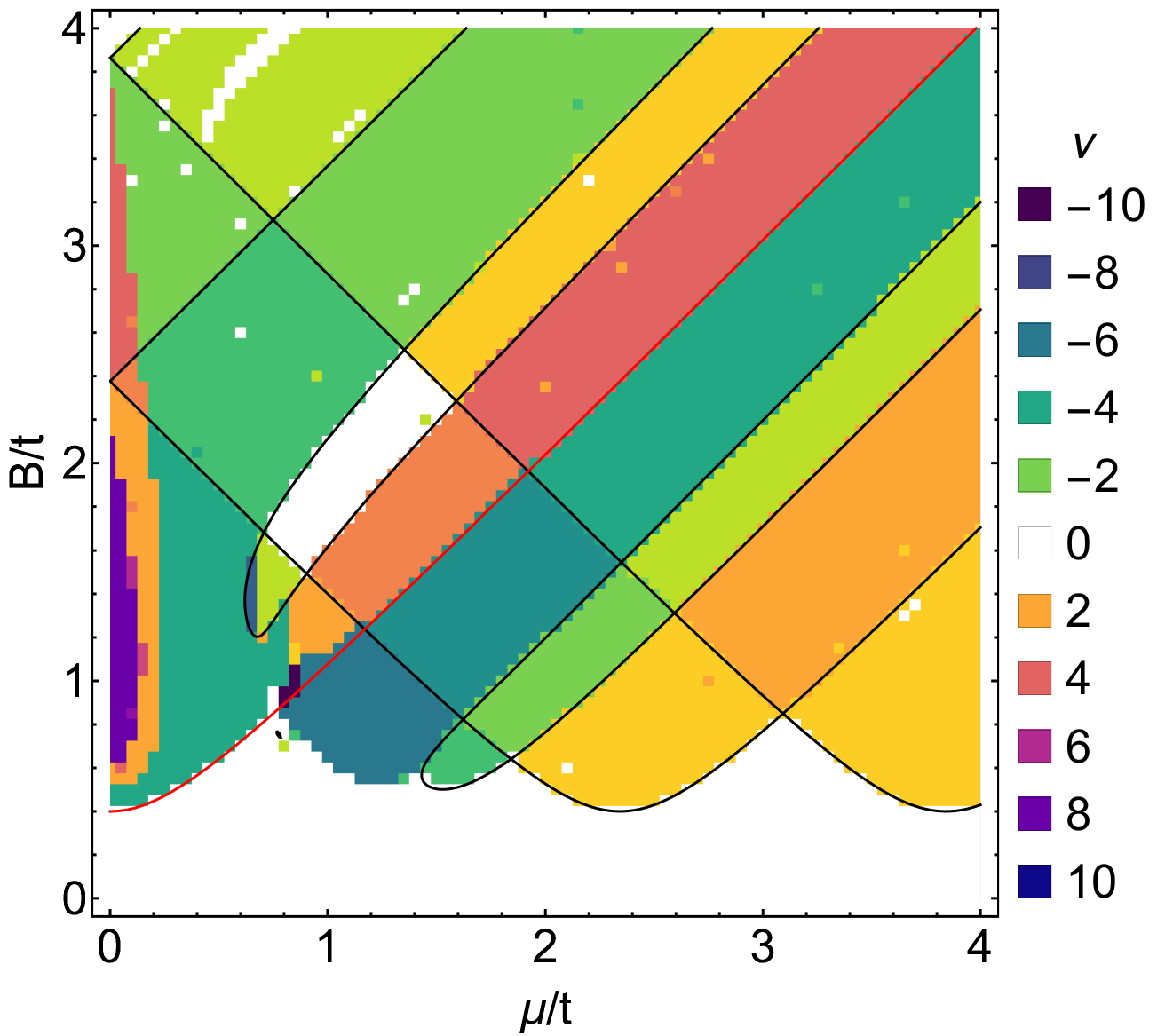}
\includegraphics[height=0.6\columnwidth]{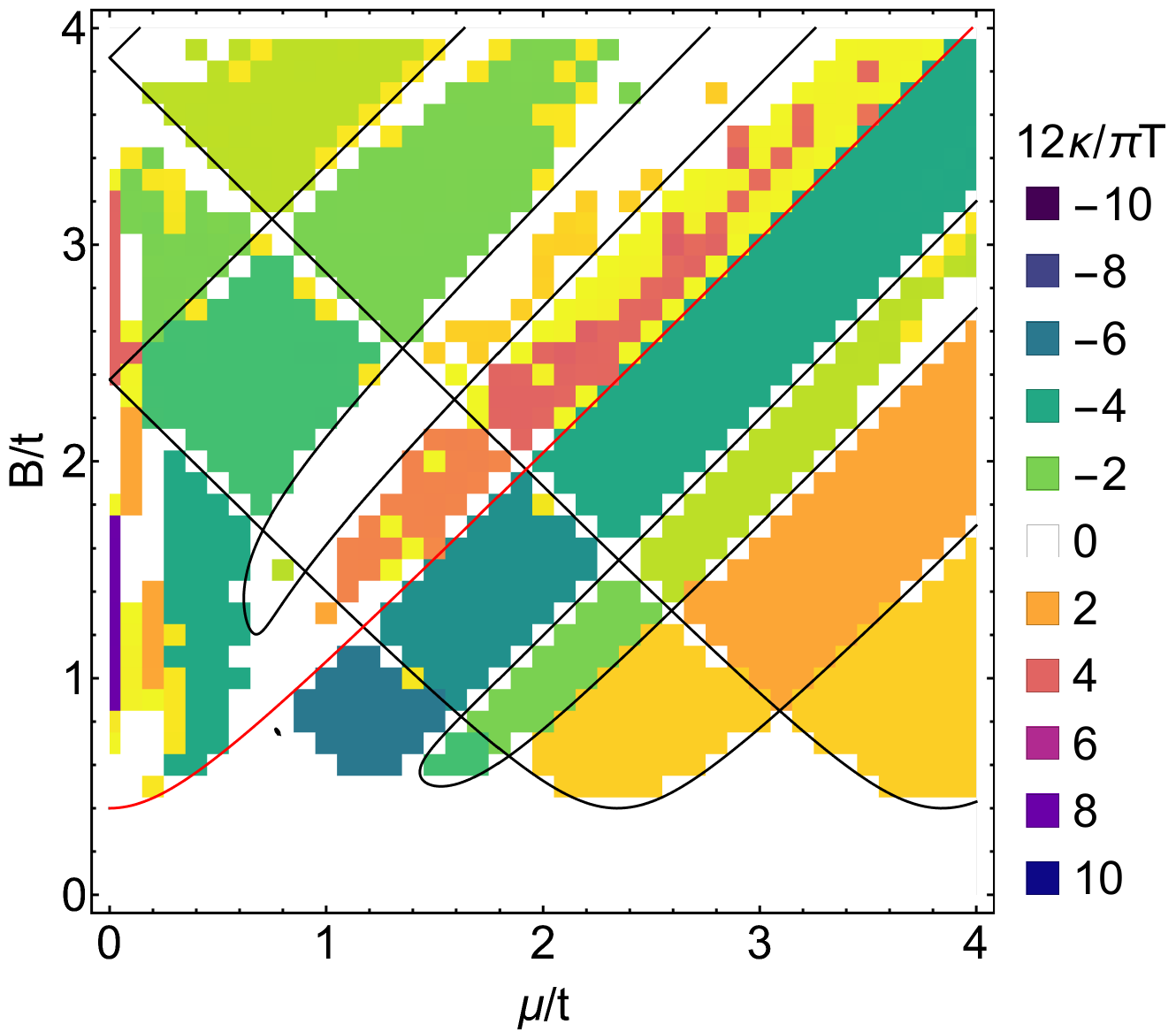}
\includegraphics[height=0.6\columnwidth]{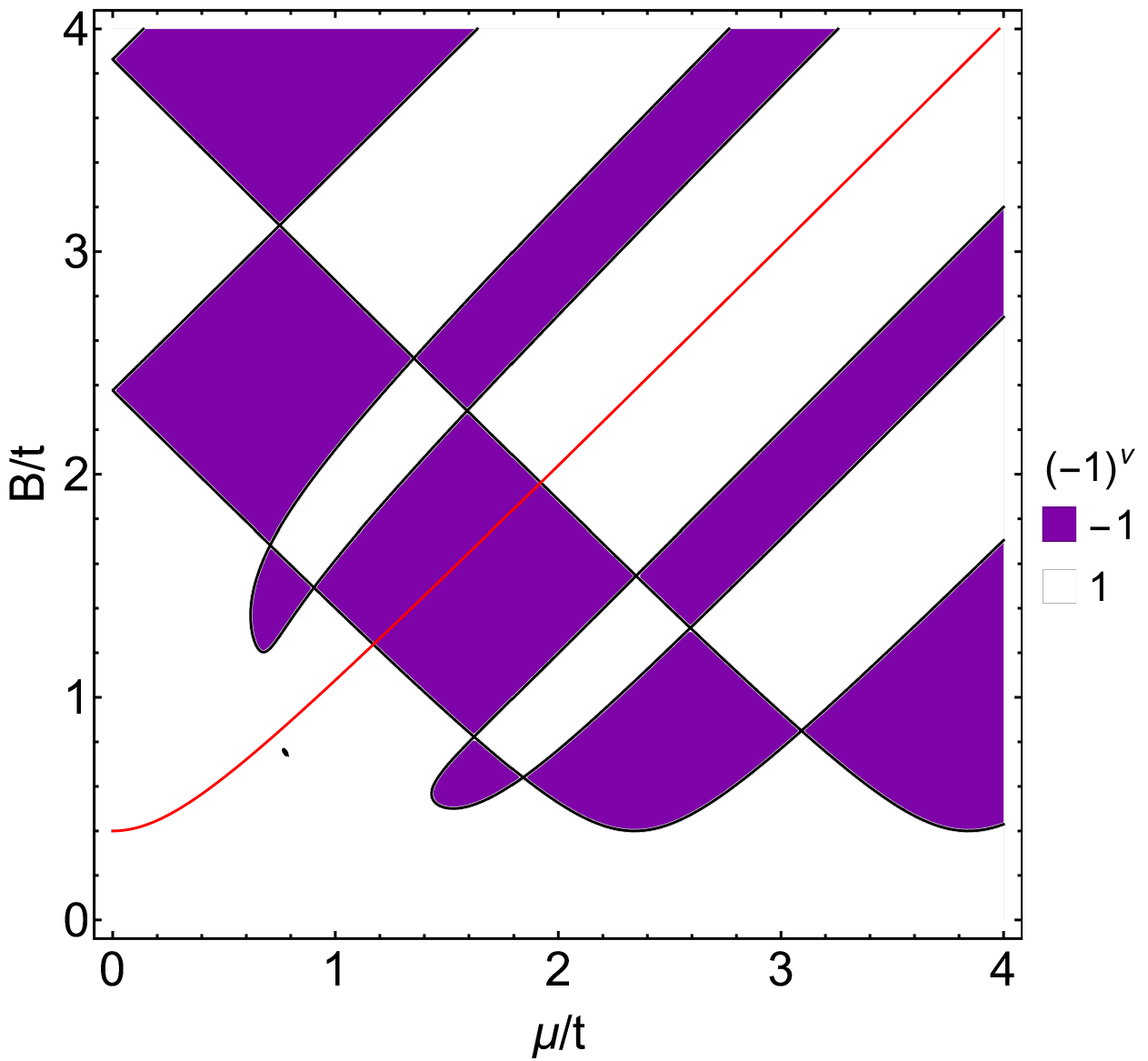}
\caption{(Color online) An exemplary topological phase diagram for the Chern number $\nu$ as a function of Zeeman field $B$ and chemical potential $\mu$ with $\Delta=0.4t$, $\alpha=0.3t$, and $t'=0.5t$. Also shown is the calculation of the quantized Hall conductance, normalized so it can be directly compared to the Chern number, and the parity of the Chern number $\delta=(-1)^\nu$. Solid black lines show the gap closings at the TRI momenta which can lead to changes in parity, and the red lines show gap closings at the Dirac points. Dashed blue lines show the parameters focused on in Fig.~\ref{fig:gap_chern}.
}
\label{fig:phase_1}
\end{figure*}

An example topological phase diagram, which contains all points in table \ref{tab:par}, is given in Fig.~\ref{fig:phase_1}. The bilayer model allows for a rich topological phase diagram with Chern numbers as large as 10, and in the following we will focus on several different phases. Further examples of phase diagrams can be found in appendix \ref{app:more_pds}.

\subsection{Parity of the Chern Number}

There is a relatively simple method for calculating the parity of the Chern number~\cite{Sato2009b} which can be generalised to hexagonal lattices~\cite{Dutreix2017,Sedlmayr2017}. It can be shown that the parity can only be changed at the TRI momenta, and hence it can be expressed in terms of quantities only at these momenta. Following a suitable transformation\cite{Sedlmayr2015,Dutreix2017,Sedlmayr2017} the Hamiltonian can be written in block diagonal form at the TRI momenta, from which the parity is extracted. Let $\tilde{\mathcal{H}}(\vec k)=\mathcal{U}^\dagger_{\vec{k}}\mathcal{H}(\vec k)\mathcal{U}_{\vec{k}}$ with the rotation
\begin{equation}
\mathcal{U}_{\vec{k}}=\frac{1-{\bm\tau}^y{\bm\sigma}^y}{2}\mathcal{A}_{\vec{k}}\mathcal{A}_{-\vec{k}}+\frac{{\bm\tau}^z{\bm\sigma}^z-{\bm\tau}^x{\bm\sigma}^x}{2}\mathcal{B}_{\vec{k}}\mathcal{B}_{-\vec{k}}\,,
\end{equation}
where
\begin{equation}
    \mathcal{A}_{\vec{k}}=\begin{pmatrix}
        e^{i\frac{k_y}{2}} & 0
        \\
        0 & e^{-i\frac{k_y}{2}}
    \end{pmatrix}\textrm{ and }
    \mathcal{B}_{\vec{k}}=\begin{pmatrix}
        0 & e^{i\frac{k_y}{2}}
        \\
        e^{-i\frac{k_y}{2}} & 0
    \end{pmatrix}.
\end{equation}
The order of matrix products here should be understood as tensor multiplication from left to right over the particle-hole, spin, AB sublattice, and finally layer, subspaces.

We then find at the TRI momenta a block diagonal Hamiltonian
\begin{equation}
\tilde{\mathcal{H}}(\vec\Gamma_i)=\begin{pmatrix}
\bar{\mathcal{H}}(\vec\Gamma_i)&0\\
0&-\bar{\mathcal{H}}(\vec\Gamma_i)
\end{pmatrix},
\end{equation}
and the parity topological invariant is simply~\cite{Sato2009b,Dutreix2017}
\begin{equation}
\delta=\left(-1\right)^\nu=\sgn\left[\det\bar{\mathcal{H}}(\vec\Gamma_1)\det\bar{\mathcal{H}}(\vec\Gamma_2)\right].
\end{equation}
Due to the symmetry of the lattice $\bar{\mathcal{H}}(\vec\Gamma_2)=\bar{\mathcal{H}}(\vec\Gamma_3)=\bar{\mathcal{H}}(\vec\Gamma_4)$ and therefore it is sufficient here to consider just two TRI momenta. When $\delta=-1$ there is a band inversion, i.e.~the parity switches between TRI momenta an odd number of times and the system is topologically non-trivial in the sense that the Chern number is odd. For $\delta=1$ the system is topologically trivial in the limited sense that the Chern number is even.

Finally, introducing $M^2=B^2-\Delta^2-\mu^2$, one finds the analytical expressions
\begin{widetext}
\begin{align}
	\det\bar{\mathcal{H}}(\vec\Gamma_1)=&
    \left[\left(M^2-9t^2\right)^2-9M^2t'^2\right]^2
    -36\mu^2\left[2t^2\left(M^2-9t^2\right)^2+t'^2\left(M^4+81t^4\right)\right]
    +1296\mu^4t^4
\end{align}
and
\begin{align}
	\det\bar{\mathcal{H}}(\vec\Gamma_2)=&
    M^8
    -2M^6 \left(-8 \alpha ^2+2 t^2+t'^2\right)\nonumber\\&
    +M^4 \left[(32 \alpha ^2 \left(3 \alpha ^2+\Delta ^2\right)+6 t^4+4 t^2 \left(t'^2-2 \left(2 \alpha ^2+\mu ^2\right)\right)+t'^4-4 t'^2 \left(4 \alpha ^2+\mu ^2\right)\right]\nonumber\\&
    -2 M^2 \bigg[2 t^6+t^4 \left(8 (\alpha^2 -\mu^2 )+t'^2\right)-8 \alpha ^2 t^2 \left(4 \alpha ^2-4 \Delta ^2-4 \mu ^2+3 t'^2\right)\nonumber\\&
    \qquad+16 \alpha ^2 \left(\left(\alpha ^2+\Delta ^2\right) \left(t'^2-8 \alpha ^2\right)+\mu ^2 t'^2\right)\bigg]\nonumber\\&
    +t^8+8 t^6 \left(2 \alpha ^2-\mu ^2\right)
    +4 t^4 \left[24 \alpha ^4+8 \alpha ^2 \left(\Delta ^2-2 \mu ^2\right)+4 \mu ^4-\mu ^2 t'^2\right]\nonumber\\&
    \qquad+32 \alpha ^2 t^2 \left[4 \left(\alpha ^2+\Delta ^2\right) \left(2 \alpha ^2-\mu ^2\right)+t'^2 \left(2 \Delta ^2+\mu ^2\right)\right]+64 \alpha ^4 \left[4 \left(\alpha ^2+\Delta ^2\right)^2-\mu ^2 t'^2\right]
\end{align}
\end{widetext}
for the determinants.

See Fig.~\ref{fig:phase_1} for an example of the parity invariant compared to the numerical calculations of the Chern number and the quantized Hall conductance. Although the parity does not contain all information about the topological phase it does include some of the most important information. It is the parity invariant which determines whether a Majorana fermion mode {\it must} be present. Additionally the parity invariant is very quick to calculate, as it is based on fully analytical expressions, in contrast to the Chern number, the numerical calculation of which can be slow for such a complicated model as the one used here.

In the following section we will show how to calculate the thermal current, from which we can find the quantized thermal Hall conductance to compare with the direct calculation of the Chern number.

\section{Quantized thermal Hall conductance}\label{sec:kappa}

The heat current for a lattice system can be found by defining a local energy term and calculating its rate of change using Heisenberg's equation of motion~\cite{Tang2019a}. Although there is no unique definition of the local energy, any reasonable definition will give the same results for the quantized edge current, and we will use a natural definition. Let us write the Hamiltonian as
\begin{equation}
	\hat{\mathcal{H}}=\sum_{j}\hat{\mathcal{H}}^{\rm os}_{j}
    +\sum_{j,i\in\mathcal{R}_j}\hat{\mathcal{H}}^{\rm bond}_{ji}\equiv\sum_j\hat{\mathcal{H}}^{\rm loc}_{j}
\end{equation}
where the index $j$ labels all sites in the two dimensional lattice and $\mathcal{R}_j$ is the set of all neighbours of site $j$. The Hamiltonian has been divided into two contributions, one defined purely onsite, and one which includes all the hopping terms. The local energy current operator is then given by Heisenberg's equation of motion~\cite{Zotos1997,Michel2006,Tang2019a}
\begin{equation}
	\hat J_j\equiv\frac{1}{2}\partial_t\hat{\mathcal{H}}^{\rm loc}_{j}= \frac{1}{4}i\left[\hat{\mathcal{H}},\hat{\mathcal{H}}^{\rm loc}_{j}\right]\,.
 \label{current_def}
\end{equation}
The numerical factors arise as we calculate the commutators using a BdG Hamiltonian. As we are interested in the currents flowing around the edges of our nanoflake we define an edge current operator
\begin{equation}
	\hat J_{\mathcal{E}}=\sum_{j\in \mathcal{E}} \hat J_j\,,
\end{equation}
where the set $\mathcal{E}$ contains all sites $j$ for which $\hat J_j$ contains terms which \emph{cross} a line extending from the centre to the edge, see appendix \ref{app:details} for an example. Thus it defines the current crossing a particular boundary. In practice we take boundaries which extend from the middle of one edge to the centre of our nanoflake.

For the expectation value we find
\begin{equation}\label{expect}
    J_{\mathcal{E}}=\sum_nf(\varepsilon_n)\langle n|\hat J_{\mathcal{E}}|n\rangle,
\end{equation}
with $f(\varepsilon_n)$ the Fermi function. The thermal Hall conductance, which is the thermal current flowing due to a temperature gradient across the sample, can be shown to be equivalent to~\cite{Kane1997}
\begin{equation}\label{kappa}
    \kappa=\frac{\partial J_\mathcal{E}}{\partial T},
\end{equation}
and the quantized thermal Hall conductance is then
\begin{equation}
    \frac{\kappa}{T}=\frac{\pi\nu}{12}.
    \label{conductance_quantum}
\end{equation}
For a non-interacting particle-hole symmetric system one can quickly verify that $\pi/12$ is the correct expression for the contribution of a single chiral edge band, see appendix \ref{app:current}.

Evaluation of the commutator Eq.~\ref{current_def}, performed using the SNEG library~\cite{Zitko2011a}, leads to the following expression for the thermal current:
\begin{align}\label{comm}
    \hat J _j=&\sum\limits_{\ell\in\mathcal{R}_j}\hat \Psi^{\dagger}_j\big[\nonumber
    \alpha\mu(\vec{d}_{j\ell}\times\vec{\bsigma})\cdot\hat{z}-i(t+t')(\mu+B\,\btau_z\bsigma_z ) \big]\hat \Psi_\ell\\\nonumber
    &+\frac{1}{2}\sum\limits_{\ell\in\mathcal{R}^2_j}\hat \Psi^{\dagger}_j\big[\alpha (t+t') (\vec{d}_{j\ell}\times\vec{\bsigma})\cdot\hat{z}-i(t^2+t'^2+t')\\
    &\qquad-\alpha^2 (\delta^x\bsigma_z\nu_{j\ell}-i\delta^y)\big]\hat \Psi_\ell\equiv \sum_\ell\hat J_{j\ell}.
\end{align}
$\mathcal{R}^2_j$ is the set of next-nearest-neighbour sites to site $j$. Here $\nu_{j\ell}$ implicitly depends on the vectors connecting the intermediate site $k$ to the sites $j$ and $\ell$ which are second nearest neighbors,  $\nu_{j\ell}=\left(\hat{d}_{jk}\times\hat{d}_{k\ell}\right)\cdot\hat z=\pm 1$, where $\hat d_{j\ell}$ is the unit vector between sites $j$ and $\ell$.

The system is solved on a finite lattice using the pybinding package~\cite{Moldovan2020}. To obtain the thermal Hall conductance we must find all terms in \eqref{comm} which cross the boundary $\mathcal{E}$, see appendix \ref{app:details}. We checked calculations for different $\mathcal{E}$ perpendicular to all four edges of our system and for different edge terminations, but find no dependence on these factors once the results have converged for large enough system sizes. In practice the sum in Eq.~\eqref{expect} is performed only for states inside the gap, which are the only ones which contribute to the heat current. Results here are for a fictitious boundary on the top zig-zag edge of the system, with the thermal Hall conductance calculated via Eq.~\eqref{kappa}.

\begin{figure}
\includegraphics*[width=0.995\columnwidth]{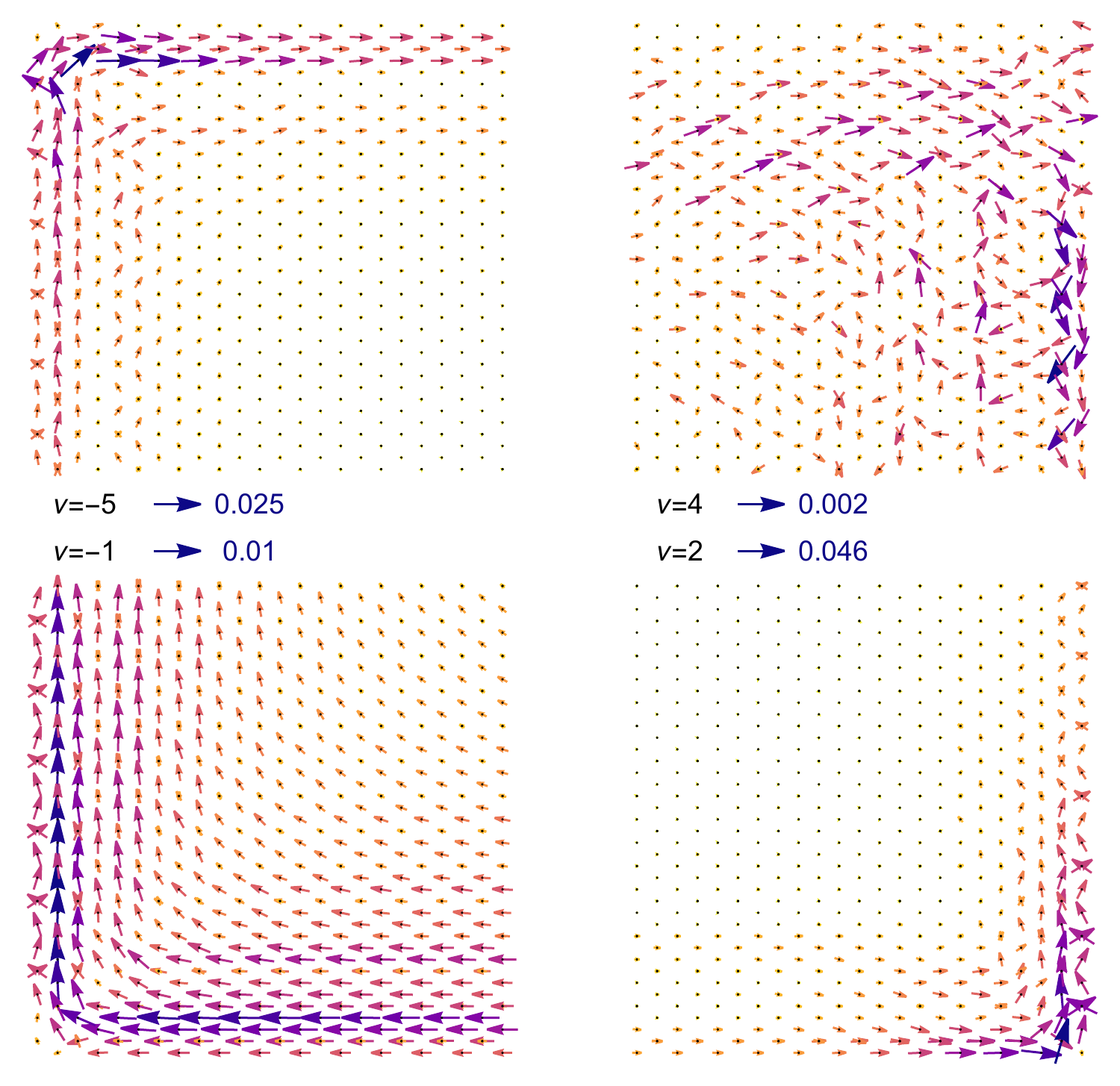}
\caption{(Color online) Examples of the heat current in four different topological phase, see Table \ref{tab:par} for parameters. Plotted is $J^{\rm p}_j$, see Eq.~\eqref{currenteq}. Each example focuses a different corner of the open nanoflake system. Although several phases follow the expected convention for the direction of flow of the heat current, we stress that it is the conductance that is quantized, not the current. The colour and size of the arrows is a guide to the magnitude of the heat current, the maximum size of $J^{\rm p}_j$ to be found in each example is given in the centre next to the arrow corresponding to this size.}
\label{fig:heat_realspace}
\end{figure}

Some examples of the real space map of the heat current are given in Fig.~\ref{fig:heat_realspace}. As $\langle\hat J_{j\ell}\rangle$ would be too complicated to show we instead plot a version of the current projected onto each site, which retains the sense of the direction of hopping. This projected current is given by
\begin{equation}\label{currenteq}
    J^{\rm p}_j=\sum_\ell \left\langle\hat J_{jl}\hat d_{j\ell}\right\rangle.
\end{equation}
Again $\hat d_{j\ell}$ denotes the unit vector between sites $j$ and $\ell$, and not an operator.

\begin{figure}
\includegraphics*[width=0.95\columnwidth]{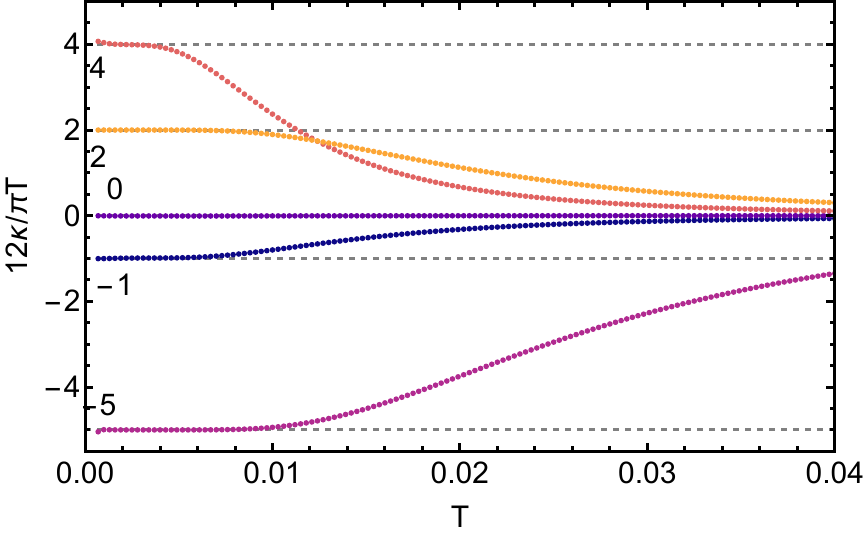}
\caption{(Color online) Thermal Hall conductance as a function of temperature $T$. See Table \ref{tab:par} for parameters for each data set. At low temperatures the plateaus can clearly be seen at the correct values in each phase.}
\label{fig:heat_plateau}
\end{figure}

To observe a correct, well-developed plateau at the value reflecting the topological invariant of the occupied bands (and the central charge of the underlying effective field theory), one has to meet certain requirements. On the one hand, the temperature has to be smaller than the effective topological energy gap, so as to not excite any bulk states and only probe the edge mode. On the other hand, the temperature has to be larger than the mean in-gap level spacing, effectively representing the velocity of the edge modes, which vastly varies among the different phases in the phase diagram. We can summarize these temperature requirements as $\Delta_{\text{eff}}< T < \Delta_{\text{eff}}/\Delta E_g$. To make sure that we can find a reasonable plateau in every identified topologically non-trivial phase, we exhaust our High Performance computing (HPC) resources, limiting the lattice to a rectangular flake of size $280a \times 262a$, where $a=1$ is the distance between the first neighbors in one honeycomb layer. This results in our finite size sample containing about $10^5$ lattice sites. This is an order of magnitude larger than previously considered in other works~\cite{Tang2019a}, allowing one to move beyond some of the limitations there and consider models with complicated internal structure and which can require larger system sizes to have well separated edge modes.

\begin{figure}
\includegraphics*[width=0.95\columnwidth]{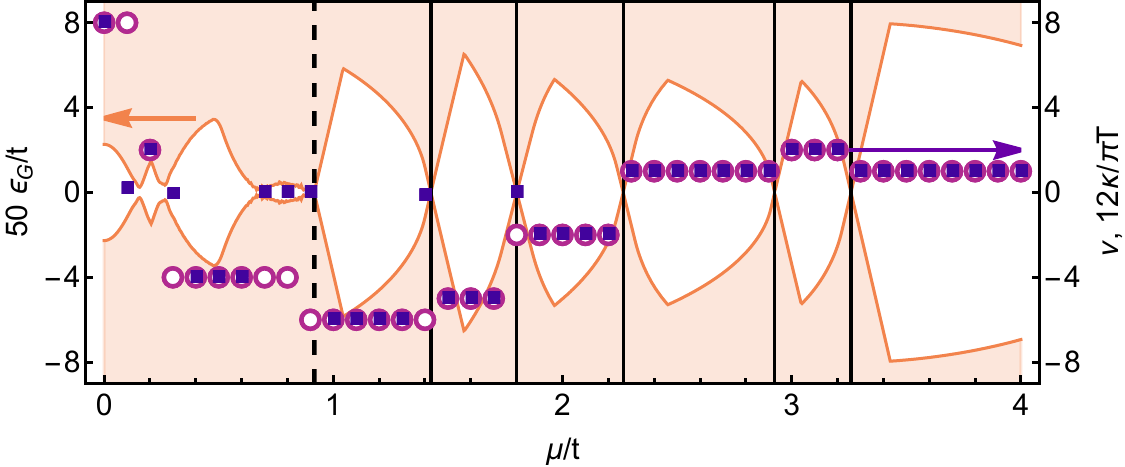}
\includegraphics*[width=0.95\columnwidth]{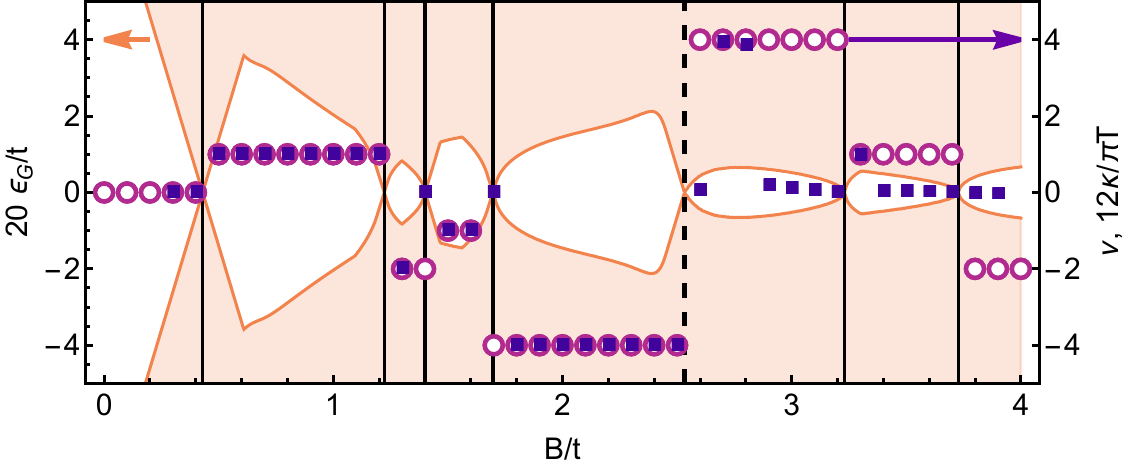}
\caption{(Color online) Chern number compared to the thermal Hall conductance in the middle of a plateau, see Fig.~\ref{fig:heat_plateau}. The size of the topological gap $\epsilon_G$ is also shown as a solid line. Results are shown for a cut through Fig.~\ref{fig:phase_1} at $\mu=2.5t$ as a function of $B$, upper panel (a), and at $B=t$ as a function of $\mu$, lower panel (b). The other parameters are $\Delta=0.4t$, $\alpha=0.3t$, and $t'=0.5t$. Large open circles show the numerically calculated $\kappa/T$ and filled squares show the Chern number. Arrows indicate the relevant axes for the data points.}
\label{fig:gap_chern}
\end{figure}

The temperature dependence of $\kappa/T$ in units of the thermal conductance quantum is presented in Fig~\ref{fig:heat_plateau}. At low temperatures, a plateau develops, whose value reflects the topological invariant, in line with Eq.~\ref{conductance_quantum}. The quality of the plateau depends strongly on the specific phase considered, and more precisely on the shape of the edge band. The examples of band structures for the probed phases are shown in Appendix~\ref{app:more_bands}. The velocity of the edge mode in the phase with $\nu=4$ is the biggest out of all probed points, and it results in the shortest plateau. We extract the value in the middle of each plateau and summarize our findings as a phase diagram in the middle panel of Fig.~\ref{fig:phase_1}. It correctly reproduces the numerical calculation of the topological invariant. For every region we are able to find points which give the correct value. The points which fail to reproduce the topological phase diagram are either due to too steep an edge band, or too small a gap. This is visible in detail in Fig.~\ref{fig:gap_chern}, where we show cuts through the phase diagram. Values found by calculating the Chern number and the thermal Hall plateau are represented by different point shapes, and the value of the gap is also visible. The thermal conductance approach performs well when there is a large enough gap in the spectrum and is seen to fail close to the phase boundaries ({\it i.e.}~the gap closings).

In an experimental situation, one cannot ignore the presence of phonons in the system, which would complicate the picture. By coupling to the edge mode the phonons start serving as a sink for the heat, and additionally the longitudinal conductance (vanishing in an ideal situation) can exceed the transverse (Hall) conductance. The low temperatures required to observe topological superconductivity can however serve as an aid. The mean free path of phonons in low temperatures is expected to be large enough to achieve ballistic transport of phonons through the bulk. While in this situation a new channel for the heat current appears and the quantization cannot be expected, a careful design of the temperature probing system and contacts can circumvent such problems~\cite{Banerjee2018a}. It is nevertheless expected that even when the coupling of the edge modes with phonons cannot be ignored, the thermal Hall conductance should be approximately quantized, as the edge contribution is only renormalized by a much smaller phonon thermal Hall term $\kappa^{ph}$~\cite{Vinkler-Aviv2018}.

\section{Conclusions}\label{sec:con}

In this article we have introduced and solved a two-dimensional topological superconductor based on a $s$-wave proximitized (111) bismuth bilayer. We first calculate the Chern number and find the topological phase diagram as a function of the various possible parameters of the model. This is then compared to the bandstructure along both zig-zag and armchair edges of the model, which can differ in details. As the numerical calculation of the Chern number is numerically costly we then demonstrate how to calculate its parity analytically, and compare this to the Chern number phase diagrams.

We then calculate the appropriate expression for the quantized thermal Hall conductance, starting with Heisenberg's equation of motion for a suitably defined local energy operator. We demonstrate that plateaus at the correct quantized values form in the $\kappa/T$ as a function of temperature in several different topological phases. We also check a topologically trivial phase which nonetheless has many trivial in-gap bands, to confirm it has zero thermal Hall conductance. Using these results as a basis we then calculate an exemplary topological phase diagram based on the quantized thermal Hall conductance to compare to the Chern number phase diagram and show that even for the complicated model used here it works reasonably well.

Here we have solved an interesting topological superconductor model which demonstrates a very rich phase diagram with large Chern numbers. Additionally these results open the way for a widespread theoretical investigation of the thermal Hall effect in many topological superconductors which could be used as the basis for experimental investigations. Furthermore the technique we use is based entirely in real space and therefore easily allows for calculations which consider the role of disorder and edge deformations, which would make interesting extensions to this work.

%%%%%%%%%%%%%%%%%%%%%%%%%%%%%%%%%%%%%
\acknowledgments

This work was supported by the National Science Centre (NCN, Poland) under the grant 2019/35/B/ST3/03625. NS thanks Cz.~Jasiukiewicz for advice with numerical integration. SG acknowledges the support of the Slovenian Research Agency (ARRS) under J1-3008. Data for the thermal Hall effect and Chern numbers can be found on Zenodo at https://doi.org/10.5281/zenodo.8420868.

%%%%%%%%%%%%%%%%%%%%%%%%%%%%%%%%%%%%%

\appendix

\section{Thermal Hall conductance for a chiral edge band}\label{app:current}

In this appendix we show the straightforward calculation for a single chiral mode in a particle-hole symmetric system. The heat current of a band with energy $\varepsilon_p$ and velocity $v_p=\partial \varepsilon_p/\partial p$ can be written as~\cite{Kane1997,Yang2020}
\begin{equation}
J_{\rm edge}=\frac{1}{2L}\sum_pv_pn_p(\varepsilon_p-\mu)\,.
\end{equation}
The distribution function $n_p$ can be taken as the Fermi function, so for a linearly dispersing mode $\varepsilon_p=-vp$ we find
\begin{equation}
J_{\rm edge}\approx\frac{1}{4\pi\hbar}\int dpv^2pf(-vp)\,,
\end{equation}
where $\mu=0$. We take the edge with the negatively dispersing band to make our sign conventions work nicely.

Now if there is a temperature gradient across a sample, with modes on each edge feeling temperatures $T\pm\delta T/2$ then
\begin{equation}
    \kappa=\frac{J_{\rm edge 1}(T+\delta T/2)+J_{\rm edge 2}(T-\delta T/2)}{\delta T}.
\end{equation}
Now as $J_{\rm edge 2}=-J_{\rm edge 1}\equiv -J_{\rm edge}$ and $f_{T+\delta T/2}(\varepsilon)-f_{T-\delta T/2}(\varepsilon)\approx \delta T\partial_T f_T(\varepsilon)$
the Thermal Hall current will be
\begin{equation}
    \frac{\kappa}{T}=\frac{1}{T}\frac{\partial J_{\rm edge}}{\partial T},
\end{equation}
and defining $x=\beta vp/2$ we find
\begin{equation}
    \frac{\kappa}{T}=\frac{k_B^2}{2\pi\hbar}\underbrace{\int\frac{x^2dx}{\cosh^2x}}_{=\pi^2/6}=\frac{\pi}{12}\frac{k_B^2}{\hbar}
\end{equation}
as expected.

\section{Additional bandstructure examples}\label{app:more_bands}

\begin{figure}
\includegraphics*[width=0.475\columnwidth]{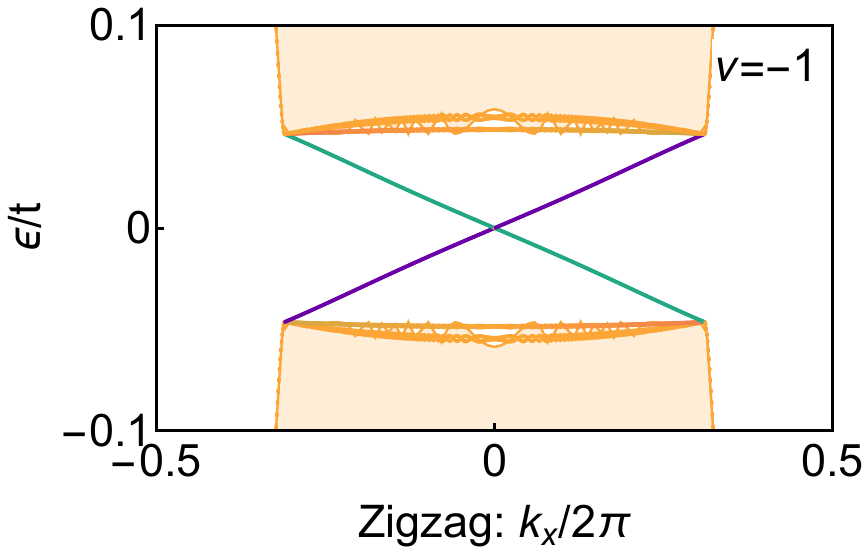}
\includegraphics*[width=0.475\columnwidth]{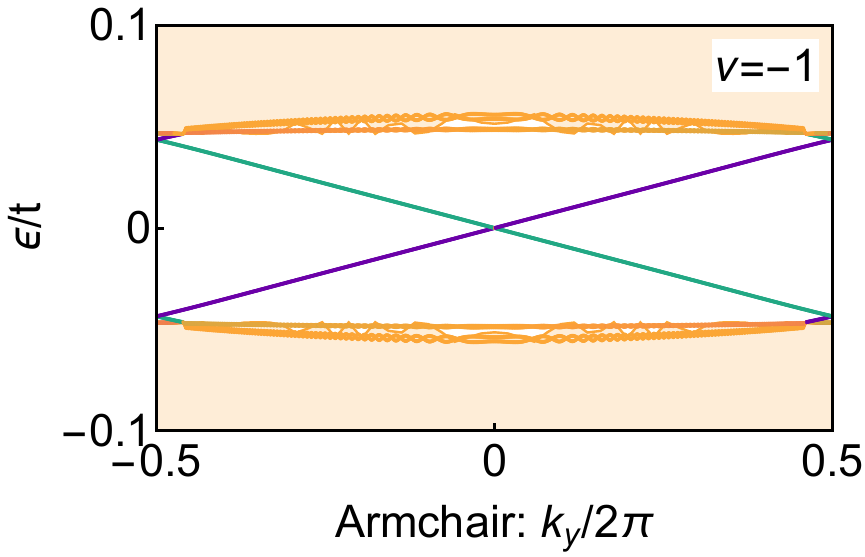}\\
\includegraphics*[width=0.475\columnwidth]{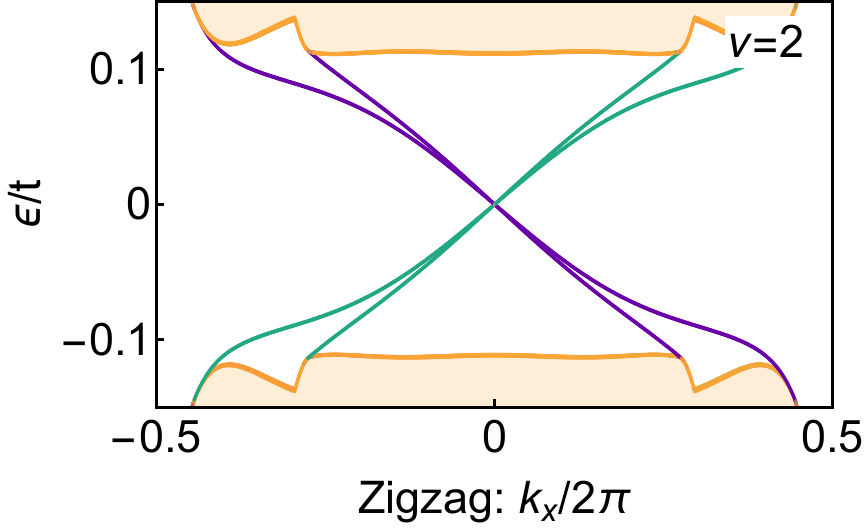}
\includegraphics*[width=0.475\columnwidth]{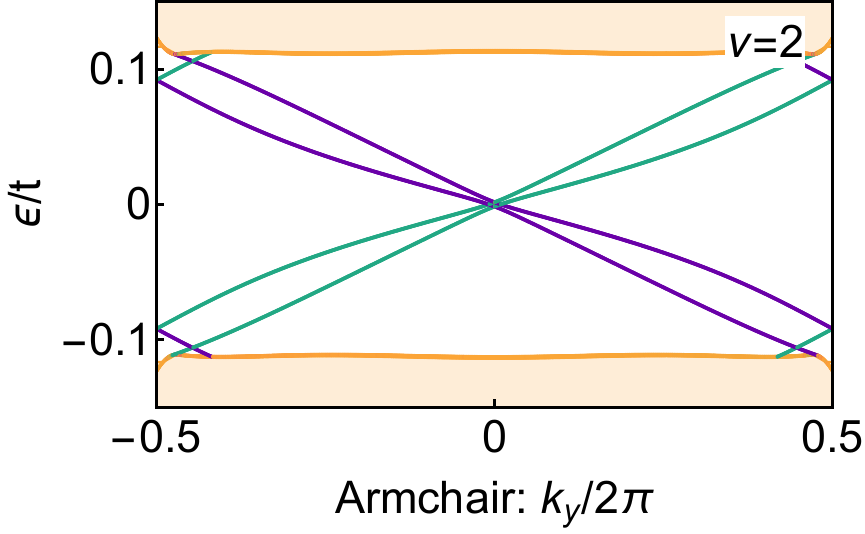}\\
\includegraphics*[width=0.475\columnwidth]{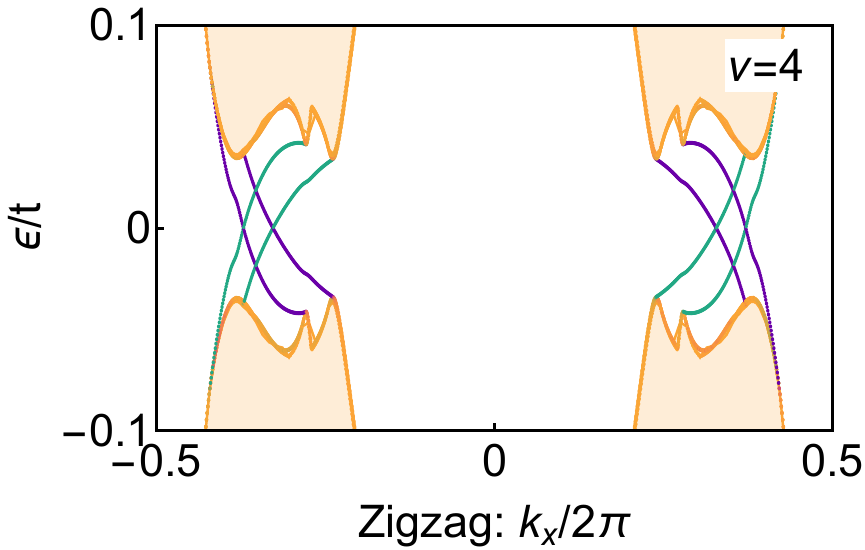}
\includegraphics*[width=0.475\columnwidth]{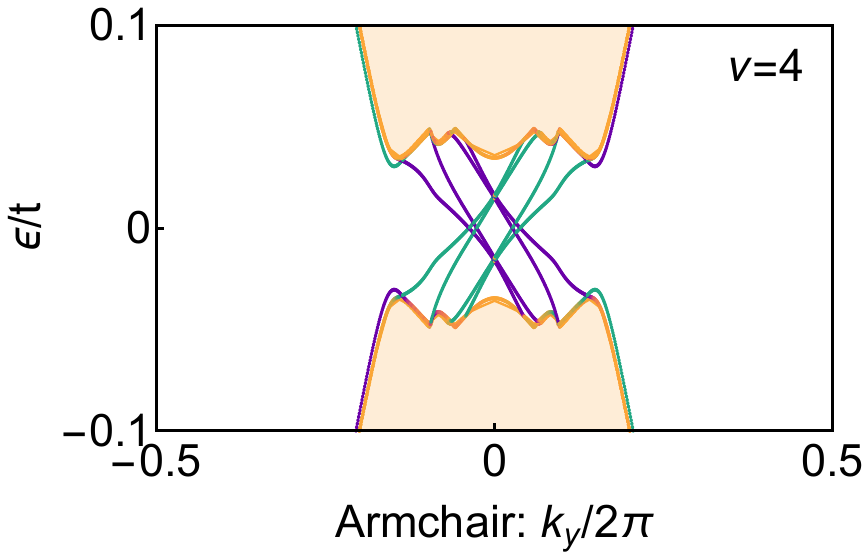}
\caption{(Color online) Band structures projected along the $x$ and $y$ directions for zig-zag and armchair edges respectively, for points in the $\nu=-1,2,4$ phases as labelled, see Fig.~\ref{fig:phase_1}. Parameters are given in table \ref{tab:par}. Light yellow shows the bulk bands, and the green and purple lines show the edge modes on the two different edges.}
\label{fig:ba_bilayer_band_exapp}
\end{figure}

In Fig.~\ref{fig:ba_bilayer_band_exapp} additional bandstructures are shown for all points at which we calculate either the local heat current, see Fig.~\ref{fig:heat_realspace}, or the thermal Hall conductance plateaus, see Fig.~\ref{fig:heat_plateau}. Edge bands along both the zig-zag and armchair edges are shown.

\section{Additional phase diagrams}\label{app:more_pds}

\begin{figure}
\includegraphics[height=0.42\columnwidth]{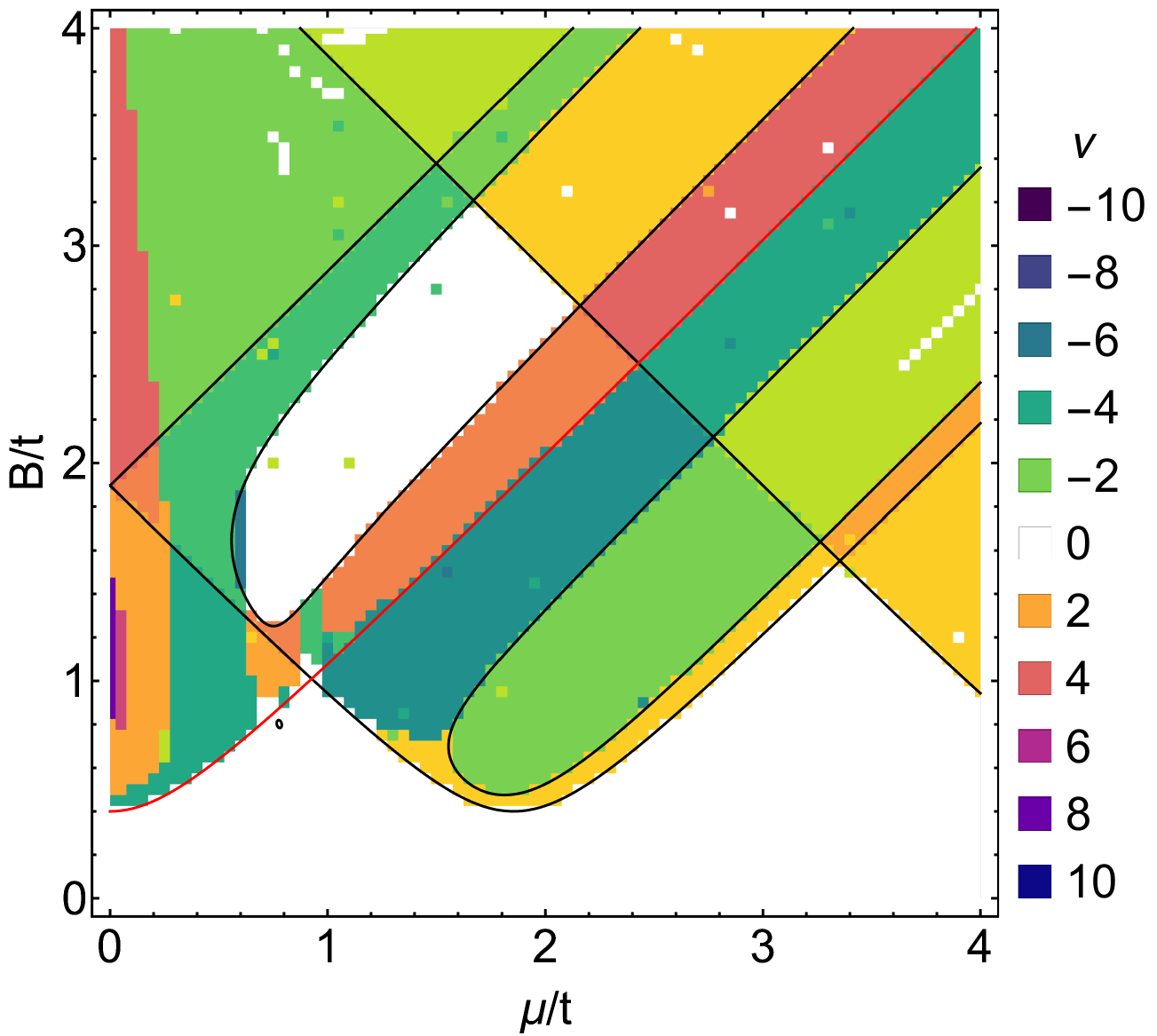}
\includegraphics[height=0.42\columnwidth]{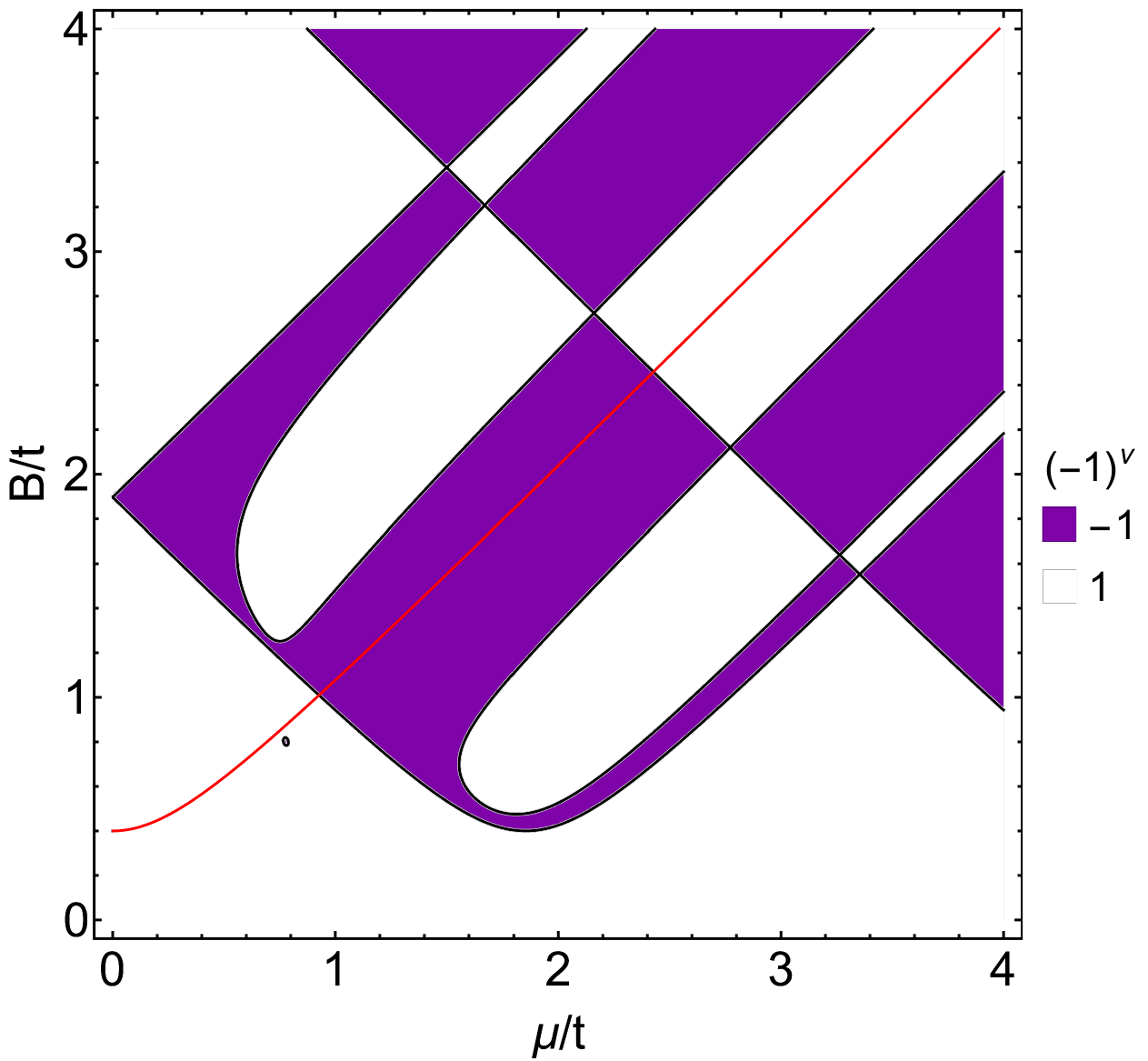}
\caption{(Color online) The topological phase diagram for the Chern number $\nu$ as a function of Zeeman field $B$ and chemical potential $\mu$ with $\Delta=0.4t$, $\alpha=0.3t$, and $t'=t$. Also shown is the parity of the Chern number $\delta=(-1)^\nu$. Solid black lines show the gap closings at the TRI momenta which can lead to changes in parity, and the red lines show gap closings at the Dirac points.
}
\label{fig:phase_2}
\end{figure}

\begin{figure}[t!]
\includegraphics[height=0.42\columnwidth]{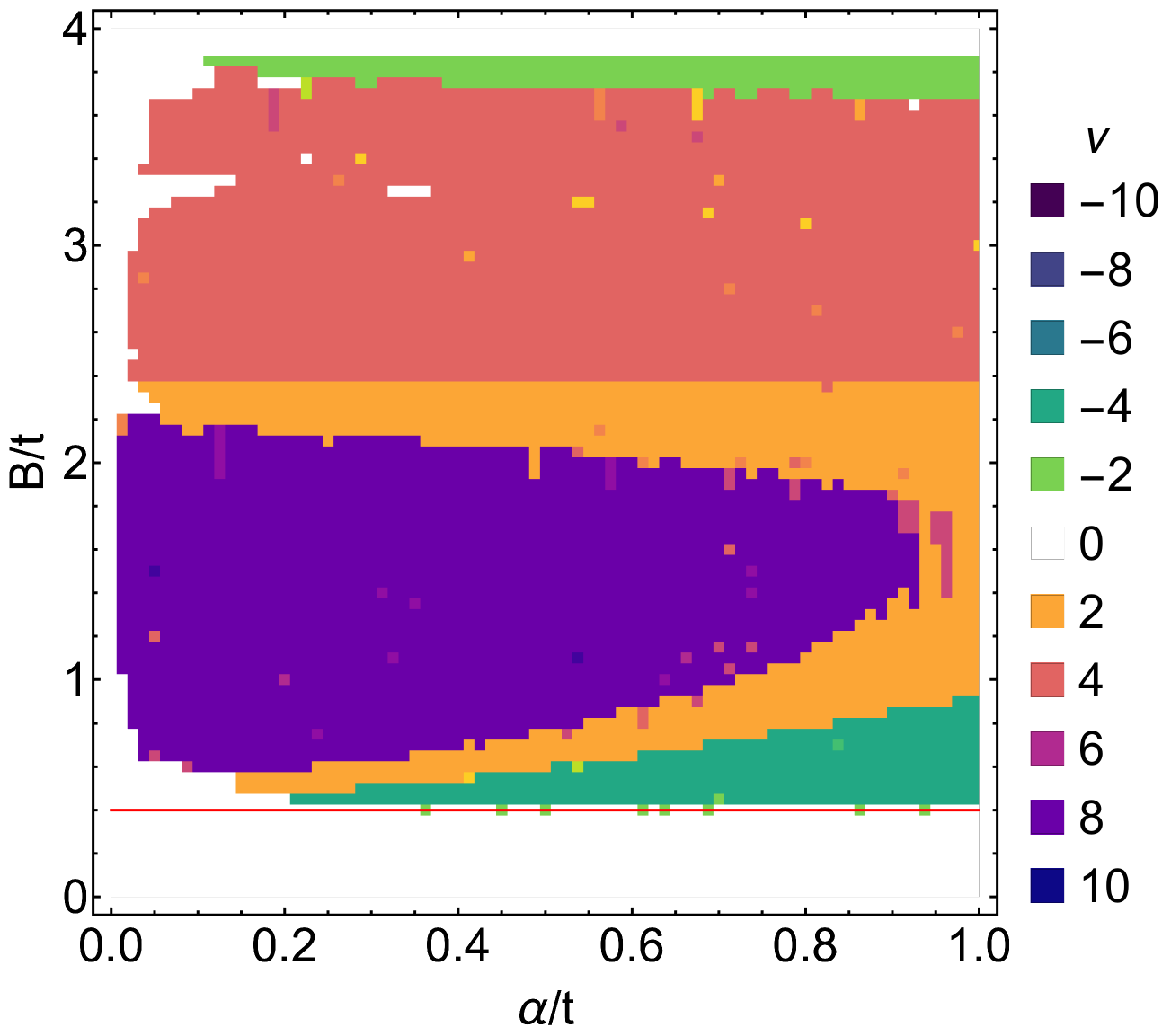}
\includegraphics[height=0.42\columnwidth]{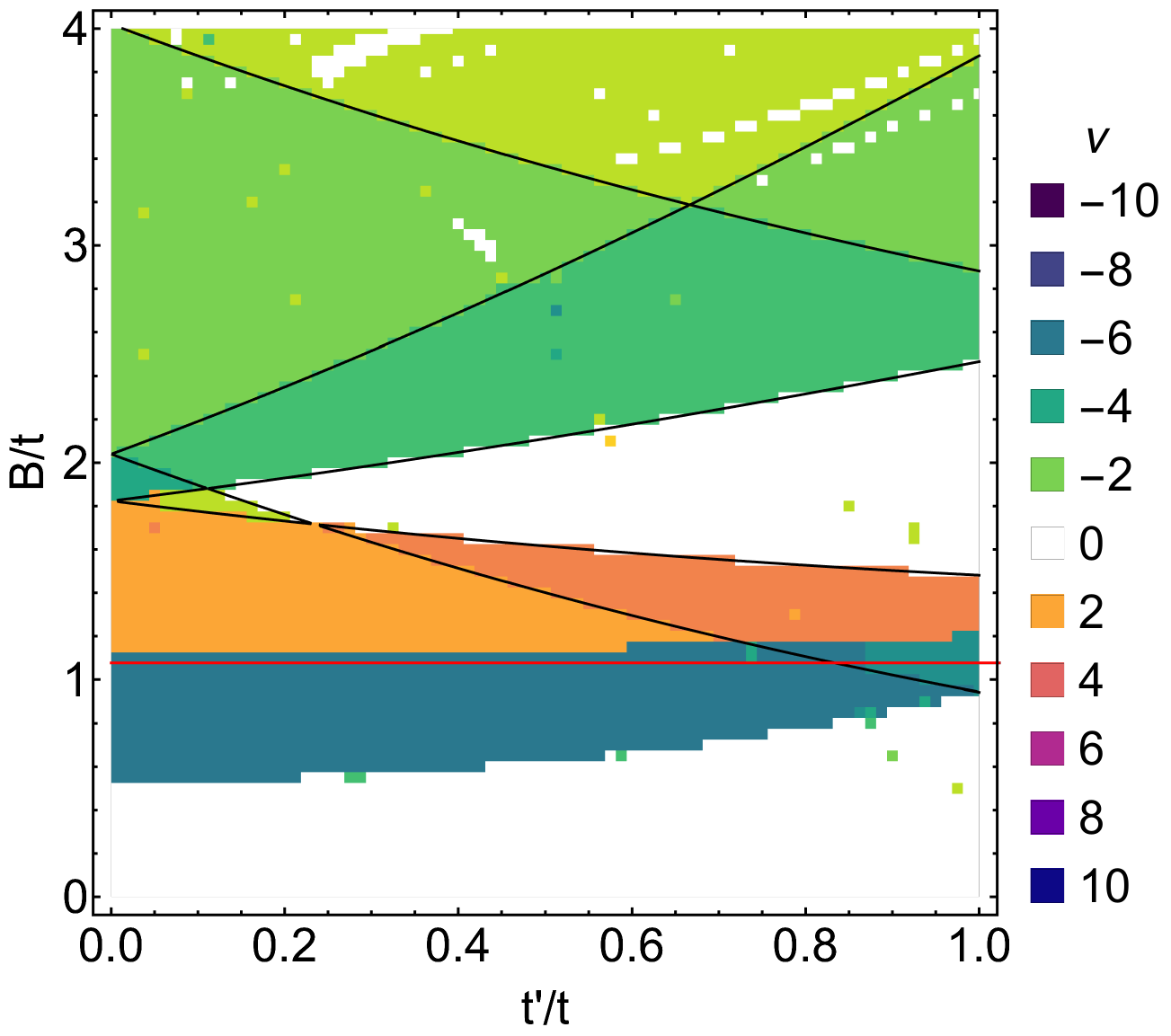}
\caption{(Color online) The topological phase diagram for the Chern number as a function of Zeeman field $B$ and either Rashba spin-orbit coupling $\alpha$ or inter layer hopping $t'$. For both panels $\Delta=0.4t$, for the left hand panel $\mu=0$ and $t'=0.5t$ while for the right hand panel $\mu=1.1t$ and $\alpha=0.3t$. Solid black lines show the gap closings at the TRI momenta which can lead to changes in parity, and the red lines show gap closings at the Dirac points.}
\label{fig:phase_3}
\end{figure}

In Figs.~\ref{fig:phase_2} and \ref{fig:phase_3} we show more examples of the Chern number and its parity. Fig.~\ref{fig:phase_2} is the same as Fig.~\ref{fig:phase_1} except the inter-layer hopping has been set to be as strong as the intra-layer hopping, which results in a similar looking phase diagram. In Fig.~\ref{fig:phase_3} we explore the dependence of the Chern number on the Rashba spin-orbit coupling and the inter-layer hopping. For changes in $\alpha$ only extremely large values substantially modify the phase diagram, and in general we find it is relatively stable as a function of $\alpha$. The inter-layer hopping has a more pronounced effect as it must interpolate between a doubled single layer system~\cite{Sedlmayr2017} and the full bilayer model. Nonetheless it is clear that for a wide range of $t'$ the topological phase diagram will show many phases.

\section{Additional details for current calculations}\label{app:details}

\begin{figure}[t!]
\includegraphics[width=0.85\columnwidth]{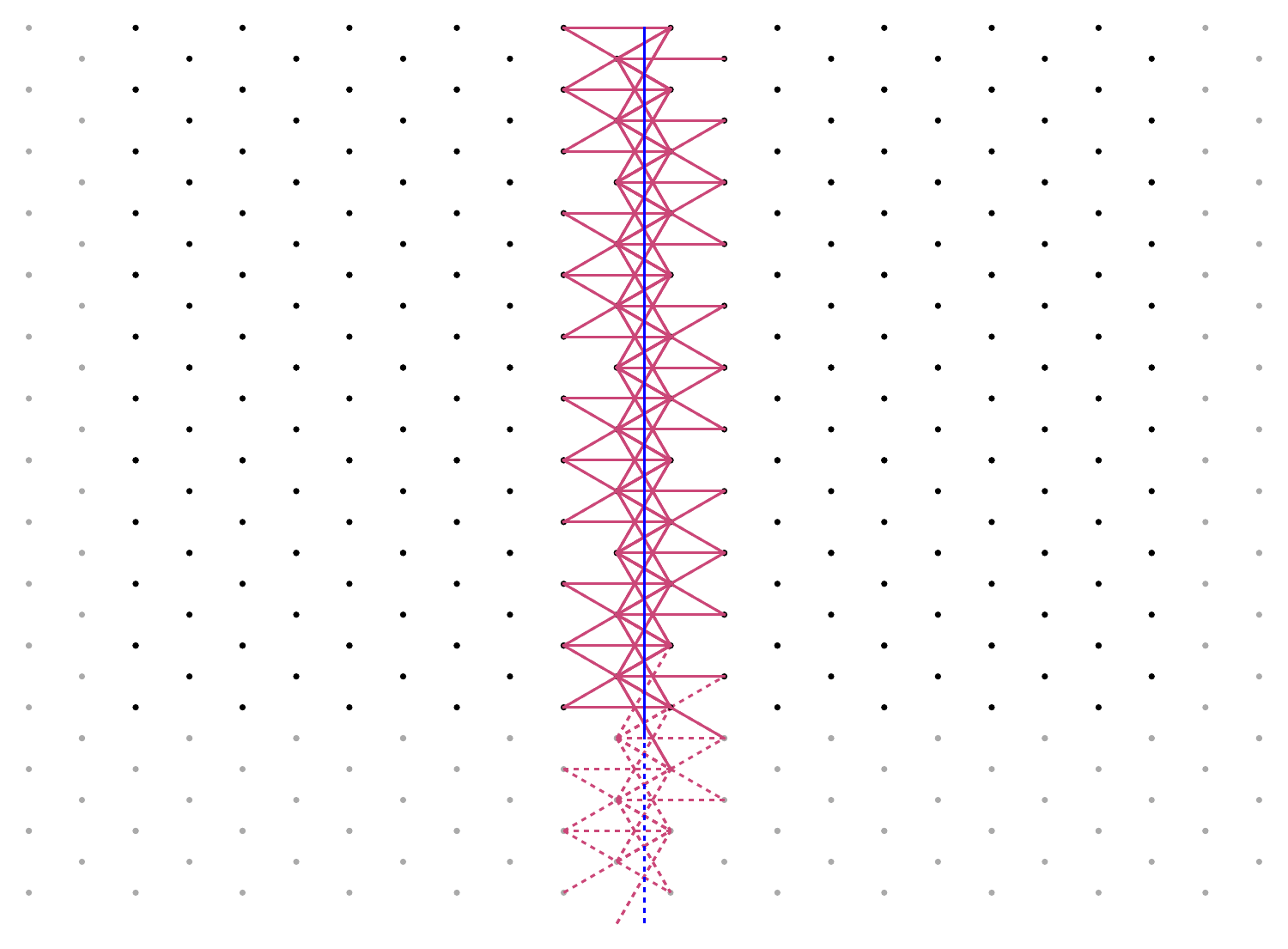}
\caption{(Color online) A top down look on the top edge of the lattice, only a small region of the whole lattice is shown with the two layers overlaid on top of each other. The blue line shows the cut across which we take all local current terms. Shown are all lattice sites connected by $\langle\hat J_{j\ell}\rangle$ across this line which form the set $\mathcal{E}$ and contribute to the calculation of the current and conductance.}
\label{fig:cut}
\end{figure}

In Fig.~\ref{fig:cut} we show the boundary across which we calculate the current on one edge, and all lattice sites coupled across this line by $\langle\hat J_{j\ell}\rangle$, i.e.~the set $\mathcal{E}$.

\begin{figure}[t!]
\includegraphics[width=0.95\columnwidth]{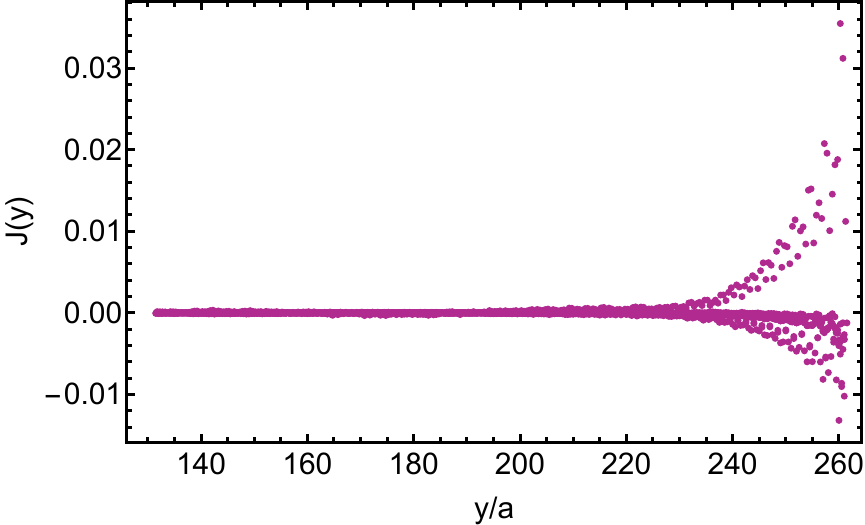}
\caption{(Color online) Spatial profile from the edge of all current terms crossing the cut $\langle\hat J_{j\ell}\rangle$, see Fig.~\ref{fig:cut}. To make $J(y)$ we take the average of the $y$-coordinates for each $\langle\hat J_{j\ell}\rangle$ which contributes. As can be seen there is no current in the bulk, in agreement with what can be seen on Fig.~\ref{fig:heat_realspace}. The edge is at $y\approx 260a$, these results are for the point in the topological phase with Chern number $-1$.}
\label{fig:profile}
\end{figure}

In Fig.~\ref{fig:profile} we given an example showing the spatial profile of the current terms which cross the boundary as a function of their distance form the edge. The terms quickly decay and there is no bulk current flowing. In this calculation both in-gap and bulk states were summed over.

%\bibliography{library}

%Control: production of eprint (1) enabled
%

\end{document}